\documentclass[%
 reprint,
nofootinbib,
 amsmath,amssymb,
 aps,
longbibliography,
]{revtex4-2}

\usepackage{graphicx}% Include figure files
\usepackage{dcolumn}% Align table columns on decimal point
\usepackage{bm}% bold math
\usepackage{hyperref}% add hypertext capabilities
\usepackage[caption=false]{subfig}% For subfigures
\usepackage{float}% to stack pictures in the same figure
\usepackage{braket}%
\usepackage{xurl}% to make URL clickable
\usepackage{cleveref}% In combination with hyperrref
\usepackage[short]{optidef}% 
\usepackage[normalem]{ulem}% strikethrough works

\usepackage{xcolor}%

\begin{document}
\preprint{APS/123-QED}

\title{Scalable embedding of parity constraints in quantum annealing hardware}% 

\author{Michele Cattelan}
    \email{Michele.Cattelan@student.uibk.ac.at}
    \affiliation{Machine Learning Research Lab, Volkswagen Group, 80805 Munich, Germany}
    \affiliation{Institute for Theoretical Physics, University of Innsbruck, Innsbruck, A-6020, Austria.}%
    
    \author{Jemma Bennett}
    \affiliation{Institute for Theoretical Physics, University of Innsbruck, Innsbruck, A-6020, Austria.}%
    
    \author{Sheir Yarkoni}
    \affiliation{Machine Learning Research Lab, Volkswagen Group, 80805 Munich, Germany}%

    \author{Wolfgang Lechner}
    \affiliation{Parity Quantum Computing GmbH, A-6020 Innsbruck, Austria}
    \affiliation{Institute for Theoretical Physics, University of Innsbruck, Innsbruck, A-6020, Austria.}%

% \date{\today}% 

\begin{abstract}
One of the main bottlenecks in solving combinatorial optimization problems with quantum annealers is the qubit connectivity in the hardware. A possible solution for larger connectivty is minor embedding. This techniques makes the geometrical properties of the combinatorial optimization problem, encoded as a Hamiltonian, match the properties of the quantum annealing hardware. The embedding itself is a hard computational problem and therefore heuristic algorithms are required. 
%These heuristics are usually time-consuming, and finding the most suitable minor embedding can be challenging, especially for large problems. Furthermore, the minor embedded Hamiltonian physical properties can differ from the original Ising Hamiltonian physical properties. This can present an issue when the properties of the original Hamiltonian are crucial in solving the optimization problem. 
In this work, we present fixed, modular and scalable embeddings that can be used to embed any combinatorial optimization problem described as an Ising Hamiltonian. These embeddings are the result of an extension of the well-known parity mapping, which has been used in the past to map higher-order Ising Hamiltonians to quadratic Hamiltonians, which are suitable for existing quantum hardware. We show how our new embeddings can be mapped to existing quantum annealers and that the embedded Hamiltonian physical properties match the original Hamiltonian properties. 

\end{abstract}

 % that is used to create a compilation of combinatorial optimization problems for different hardware topologies

% By employing this mapping, no further embedding techniques are required and the time required to initialize the problem in the quantum annealer is nullified

\maketitle

%\tableofcontents

\section{Introduction}\label{sec: intro}%

Quantum annealing (QA) consists of a family of metaheuristic optimization algorithms that use quantum fluctuations to solve combinatorial optimization problems. A system of qubits is used to implement the variables of the combinatorial optimization problem, to which a continuous quantum evolution is applied. The first proposals of using QA to solve hard problems involved simulating transverse-field Ising Hamiltonians, where a simple initial Hamiltonian would be slowly transformed to a classical spin-glass representation of the problem of interest~\cite{kadowaki1998quantum}. Since then, different paradigms have been proposed to use QA in practice~\cite{Hauke2020}. Special-purpose QA hardware, provided by D-Wave Systems, has been built to implement QA and has been demonstrated to use quantum entanglement to simulate such Ising Hamiltonians~\cite{johnson2011quantum,lanting2014}. Despite some works that demonstrated its potential~\cite{king2019quantum, denchev2016computational}, to date, there are no practical advantages to using this technology with respect to classical solutions. There are several bottlenecks in the application of quantum annealing to solve combinatorial optimization problems~\cite{yarkoni2022quantum}. 

One of the most important considerations of using quantum annealing hardware in practice is the fixed connectivity (or topology) of the layout of qubits and connections between them. We can describe the connectivity of the quantum processor as a graph where the qubits of the quantum annealer are nodes and the couplers between them are the edges. Therefore, to solve arbitrary problems, we must map these problems formulated as Ising Hamiltonians to match the connectivity of the quantum annealer. In other words, the problem graph must be a subgraph of the quantum annealer hardware graph. This mapping process is called minor embedding and different techniques have been developed to solve it~\cite{boothbyclique}. The minor embedding problem is known to be hard~\cite{schoning1988graph} and so no polynomial-time algorithm is known to solve it optimally. Hence, classical heuristics and meta-heuristics to solve the minor embedding problem are required. The time needed by these algorithms to find a suitable minor embedding of the problem Hamiltonian is not negligible and it represents one of the main bottlenecks for QA to become practical and compete against classical optimization solvers~\cite{yarkoni2018}.

One proposed method to avoid the minor embedding issue is to reformulate the problem using the so-called ``parity mapping''--- here, each qubit in the Hamiltonian no longer represents a single variable in the optimization problem, but rather each qubit now represents the relative orientation of multiple spins in the Ising Hamiltonian~\cite{ender2023parity}. This method uses specific sets of constraints to encode arbitrary Ising Hamiltonians using local multi-body interactions. A specific case of such a parity mapping, known as the LHZ transformation, yields an all-to-all connected Hamiltonian~\cite{lechner2015quantum}.

In this work, we focus on an extension of the parity transformation to map parity Ising Hamiltonians to existing quantum annealers with at most quadratic interaction terms. We achieve this using constraints with at most quadratic interaction terms to replace the multi-body terms in the original formulation of the parity mapping. We specifically show how to map these constraints in a generalizable way to existing D-Wave quantum annealers and avoid the time-consuming minor embedding step. 
Moreover, we investigate the relationship between the different formulations and the minimum spectral gap of these Hamiltonians, a crucial element in understanding the underlying physics of quantum annealing. 

Finally, we compare between numerical simulations and experiments on a D-Wave QPU to understand if current quantum annealers can simulate the expected physical properties of the studied Hamiltonians. 

The paper is structured as follows: in \cref{sec: QA} we introduce quantum annealing and its related uses and bottlenecks; in \cref{sec: parity} we briefly explain what the parity mapping is and its most direct implementation (the LHZ transformation); then, in \cref{sec: parity on dwave} we present how to extend the parity mapping and how to use it to describe novel scalable embeddings on quantum hardware; then, in \cref{sec: results} we present the results of both numerical simulations and data from D-Wave hardware.

\section{Quantum annealing (QA)}\label{sec: QA}
Quantum annealing can be viewed as a relaxation of Adiabatic Quantum Computing (AQC)~\cite{farhi2000quantum}. AQC is a form of computing that relies on the quantum adiabatic theorem~\cite{kato1950adiabatic} which states that, if a closed quantum system is prepared in an eigenstate, it remains in this eigenstate if the system is evolved ``slowly'' enough. However, QA is implemented on programmable quantum hardware, hence, it is realized in open quantum systems. Moreover, the evolution that describes QA does not necessarily fulfill the conditions of the adiabatic theorem and thus QA can be seen as a relaxation of AQC.

QA can be described by the following steps. First, we prepare a quantum system in the ground state of a known, simple initial Hamiltonian, $H_i$. Next, we describe a Hamiltonian $H_f$, which encodes in its ground state the desired outcome of the computation. (We consider the case where the ground state is the optimal solution of a combinatorial optimization problem.) Then, the system is evolved according to a so-called annealing schedule, which changes the Hamiltonian of the system from $H_i$ to $H_f$. Eventually, the result of the computation is read by measuring the quantum system.

We can write a time-dependent Hamiltonian that describes the evolution from the ground state of $H_i$ to the ground state of $H_f$ as:
\begin{equation}\label{eq: AQC}
    H(t) = A(t)H_i + B(t)H_f,
\end{equation}
where $A(0)=B(T)=1$ and $A(T)=B(0)=0$.

Typically, the initial Hamiltonian $H_i$ is defined using $\sigma_x$ Pauli operators, and the final Hamiltonian $H_f$, which encodes the combinatorial optimization problem, is defined by using $\sigma_z$ Pauli operators. We describe these Hamiltonians as 
\begin{align}\label{eq: Hamiltonian form}
    H_i & = \sum_{i=0}^{n-1} \sigma_x^i\\
    H_f & = \sum_{i=0}^{n-1} h_i\sigma_z^i + \sum_{i<j}J_{ij}\sigma_z^i\sigma_z^j,
\end{align}
with programmable $h_i, J_{ij}\in\mathbb{R}$ and $\sigma_{*}^i$ is the $*$-Pauli operator that acts on the $i$-th spin, where $*$ is either $z$ or $x$. Furthermore, we can define the functions $A(t)=\left(1-s(t)\right)$ and $B(t)=s(t)$ such that they depend on a single function $s:[0,T]\mapsto[0, 1]$, which defines the annealing schedule, and $s$ is referred to as normalized time. Hence, the time-dependent Hamiltonian becomes
\begin{equation}\label{eq: QA}
    H(t) = \left(1-s(t)\right)H_i + s(t)H_f.
\end{equation}
To solve an optimization problem with QA, we first prepare the system in the ground state of $H_i$. Typically, in the case of choosing $H_i$ such as in \eqref{eq: Hamiltonian form}, the ground state is the superposition of all the possible computational basis states; after initial preparation, we evolve the system according to $H(t)$. At the end of the evolution the system is measured in the computational basis.

As mentioned previously, because QA hardware is an open quantum system, the adiabatic theorem in general does not hold. Therefore the condition of ``slow'' evolution required for AQC is not well described. Thus, we obtain a relaxation of adiabatic evolution, which is quantum annealing. Here, instead of guaranteeing a smooth transition to global optima of $H_f$, we obtain a bound on the non-zero probability of observing ground states.  A parameter important to QA is the minimum difference between the two lowest instantaneous eigenvalues of $H(t)$, also known as the minimum spectral gap. The minimum spectral gap affects the total time $T$ needed to perform an evolution. Previous results show that in order to satisfy the adiabatic condition, T must scale with the minimum spectral gap. We denote this minimum as $\Delta_{0,1}\coloneqq\min_{s\in[0,1]}\epsilon_1(s) - \epsilon_0(s)$, where $\epsilon_i(s)$ are the instantaneous eigenvalues of the Hamiltonian $H(s)$ (recalling that $s$ is normalized time). The adiabatic condition can be written as 
$$T\in\Omega\left(\frac{1}{\Delta_{0,1}^2}\right),$$
meaning that the runtime of QA depends inversely on how the spectral gap of the Hamiltonian $H(s)$ closes~\cite{Hauke2020}.
Even though the guarantee of the adiabatic theorem does not apply to QA, the minimum spectral gap of $H(t)$ influences its performance. 
As noted in~\cite{dickson2013thermally}, even with a small $\Delta_{0,1}$ and with $T$ that does not fulfill the adiabatic theorem, QA performance can still match the expected theoretical performance of adiabatic evolution. Thus, the analysis of the minimum spectral gap is crucial to studying the performance of QA. 

Current state-of-the-art quantum annealers, and specifically those provided by D-Wave Systems, use superconducting circuits to implement qubits. Qubit pairs are connected via couplers, with a fixed connectivity between them typically referred to as the \emph{hardware graph}. Usually, not all the qubits are connected to each other. Thus, if we consider the qubits as nodes and the couplers connecting them as edges, we can describe the quantum annealer as a graph. Since the couplers are physical connections between qubits, the topology of the graph described by the quantum hardware is fixed. Such fixed connectivity of the qubits means that not all Ising Hamiltonians are native to the hardware graph. One solution to this problem is to map the spins in the Hamiltonian $H_f$ to more than one physical spin--- called a chain--- thus increasing the degree of the spin. The problem of finding such groups of qubits to implement the problem Hamiltonian can be translated to finding the minor of the graph representing the problem that is a subgraph of the quantum annealer graph. In this context, this problem is known as the minor embedding. In general, finding a suitable minor embedding is hard~\cite{schoning1988graph} and time-consuming~\cite{yarkoni2018}. 

After finding the suitable minor, the problem Hamiltonian is modified accordingly so that the ground state of the minor embedded Hamiltonian matches that of the original (logical) problem Hamiltonian. To ensure all spins in each chain are in the same spin state in the ground state of $H_f$, we add an additional ferromagnetic coupling term between all qubits in the same chain to the Hamiltonian.  
After quantum annealing is executed, each spin is measured in the computational basis. Then, the results are stored and every chain is checked. When a chain does not contain spins with the same value, it must be corrected by choosing a single spin state--- we call such a chain ``broken''. To correct this we can select as the spin state of the chain the state that appears the most in the set of spins considered in the chain (known as majority voting). The process of choosing the proper state value of a chain is called ``fixing''. After fixing all chains we can read a solution of the implemented Ising Hamiltonian. Thus, arbitrary combinatorial optimization problems can be solved using fixed-topology quantum annealers.

\section{Parity quantum computing (PQC)}\label{sec: parity}
In this section, we present the parity mapping and explain how an Ising Hamiltonian with terms of any order can be transformed and simulated using a transverse-field Ising Hamiltonian. 

\subsection{LHZ triangle}
The first proposal of the parity mapping is the so-called Lechner-Hauke-Zoller (LHZ) triangle~\cite{lechner2015quantum}\footnote{In this paragraph, we consider only Ising Hamiltonian with terms of at most quadratic order. A generalization of the LHZ triangle for higher-order interactions is proposed in the original publication~\cite{lechner2015quantum}}. Consider an Ising Hamiltonian of the following form
\begin{equation}\label{eq: Ising}
    H=\sum_i h_i\sigma^i_z + \sum_{i<j} J_{i, j}\sigma^i_z\sigma^j_z.
\end{equation}
We can replace every term that appears in \cref{eq: Ising} with a new spin:
\begin{align*}
    \sigma_z^i & \mapsto\tilde{\sigma}_z^{(i)},\\
    \sigma_z^i\sigma_z^j & \mapsto\tilde{\sigma}_z^{(i, j)}.
\end{align*}
Hence, we can rewrite \cref{eq: Ising} as
\begin{equation}\label{eq: local field parity}
    H_l = \sum_i h_i\tilde{\sigma}_z^{(i)} + \sum_{i<j} J_{i, j}\tilde{\sigma}_z^{(i, j)}.
\end{equation}
We refer to the spins of $H$ as logical spins, whereas we refer to the spins in $H_l$ as physical spins or parity qubits. Notice that the number of physical spins used in $H_l$ is larger than the number of spins in $H$. This implies that the Hilbert space defined by $H_l$ is larger than the Hilbert space of the logical Hamiltonian. 
Therefore, due to the increased degrees of freedom, there exists solutions in the Hilbert space of the physical spins that do not represent any solution in the logical Hilbert space--- we denote these as invalid solutions. Hence, we must add penalty terms ($H_P$) to $H_l$ to restrict the ground state distribution of the physical Hamiltonian so it matches the ground state distribution of the logical Hamiltonian $H$. Therefore, we define the physical Hamiltonian $\tilde{H} = H_l + \lambda H_P$, with a properly tuned $\lambda$ so that only valid solutions are in the ground state. 

Notice that if $H_P$ is not chosen properly, then the parity of the terms of the original Hamiltonian $H$ is not preserved. This can be understood by looking at subsets of qubits that interact. For example, if we have four qubits that interact pairwise, with interactions $\sigma_z^i\sigma_z^j,\; \sigma_z^j\sigma_z^k,\; \sigma_z^k\sigma_z^p$ and $\sigma_z^p\sigma_z^i$, we can see that
\begin{equation}\label{eq: loop condition}
    \sigma_z^i \sigma_z^j \sigma_z^j \sigma_z^k \sigma_z^k \sigma_z^p \sigma_z^p \sigma_z^i = 1.
\end{equation}
This equation holds for all the possible choices of indices $i, j, k$, and $p$, as they appear exactly twice in the multiplication and $\left(\sigma_z\right)^2=\mathbb{I}$. 

This condition is not fulfilled anymore if we substitute the interactions with the physical spins. Therefore, if we replace the logical spins in \cref{eq: loop condition} with physical spins we must ensure that
\begin{equation}\label{eq: parity constraint condition}
    \tilde{\sigma}_z^{(i, j)}\tilde{\sigma}_z^{(j, k)} \tilde{\sigma}_z^{(k, p)} \tilde{\sigma}_z^{(p, i)} = 1 
\end{equation}
for all $i, j, k$ and $p$. 
We can solve this for any choice of indices by introducing an additional constraint (which is called the parity constraint) to guarantee that only valid spin configurations appear in the ground state. 
However, there are factorially many such constraints and including them all leads to redundancy. Therefore, we must find at least the minimum number of such parity constraints, which is known to be exactly $K - N$~\cite{lechner2015quantum}, where $K$ is the number of physical spins and $N$ is the number of logical spins. 
The easiest way to find these constraints is to consider the graph $G_H$ that describes the logical Hamiltonian $H$--- we consider each spin as a node and draw an edge between nodes that represent interacting spins in the Hamiltonian. We can define $H_P$ by identifying at least $K-N$ loops of length $4$ and $3$ in $G_H$ such that every node appears in at least one loop~\cite{ender2023parity}. 
To ensure the system is properly constrained, every parity qubit in $H_l$ must be contained in at least one parity constraint, also known as a plaquette. The parity Hamiltonian can then be defined as 
\begin{equation}\label{eq: parity constraint ham}
    H_P = - \sum_{l=0}^{K-N-1} \tilde{\sigma}_z^{l, n}\tilde{\sigma}_z^{l,e} \tilde{\sigma}_z^{l, w} \left(\tilde{\sigma}_z^{l,s}\right),\footnote{The term in brackets is used only if the loop considered has length $4$.}
\end{equation}
where $l$ identifies the plaquette, and indices $n, e, w$ and $s$ identify the physical spins in plaquette $l$. We can easily see that the minimum of the Hamiltonian $H_P$ as defined in \cref{eq: parity constraint ham} is obtained when all terms equal $-1$, which is equivalent to the condition described in \cref{eq: parity constraint condition}. 

\begin{figure}
    \centering
    \includegraphics[scale=0.17]{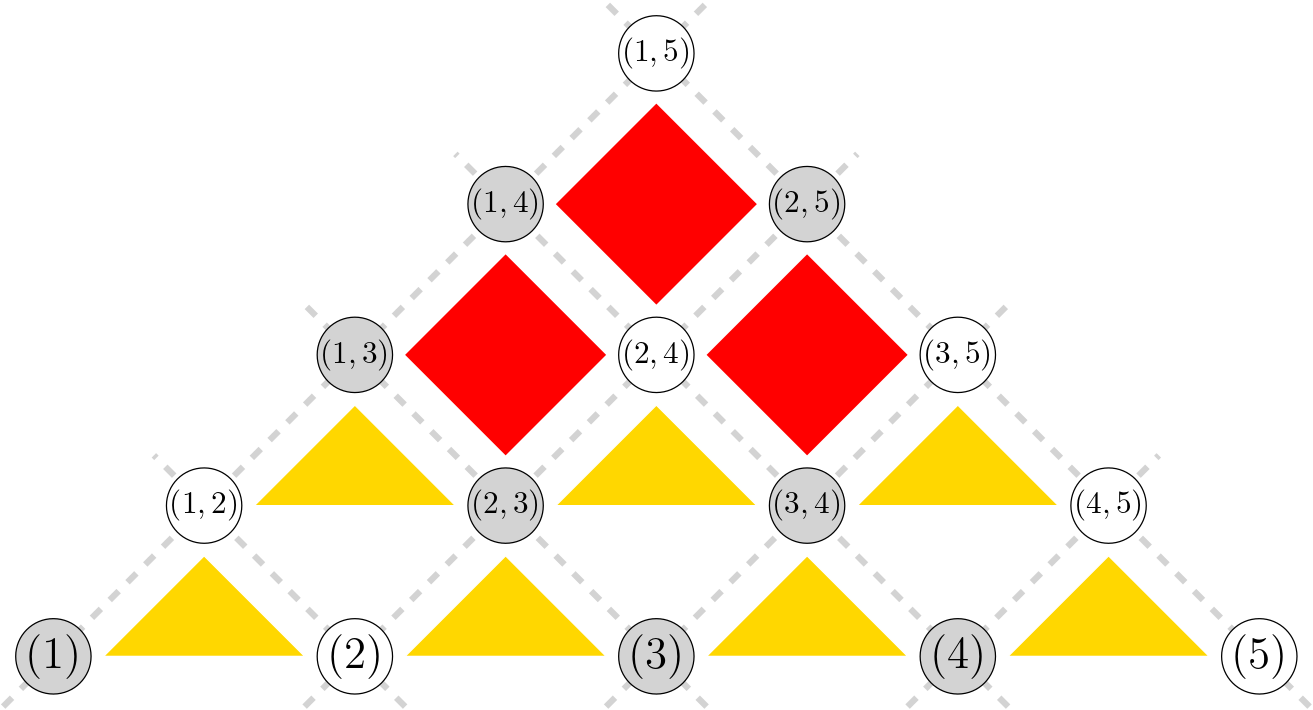}
    \caption{Visual representation of the parity Hamiltonian $\tilde{H}=H_l + \lambda H_P$, as described in \cref{eq: local field parity} and \cref{eq: parity constraint ham}. The nodes represent the physical spins used to simulate the logical Ising Hamiltonian. Their labels refer to the indices of the parity qubits. The triangles and the squares describe the multi-body interactions in $H_P$, that are needed to ensure the validity of the solutions according to the condition in \cref{eq: parity constraint condition}. We use the dashed grey lines to show how the indices are shared among the parity qubits. Notice that each qubit on the same line shares one index. The local field of the physical parity qubit representing an unused interaction is tuned to $0$. As an example in the above figure, the grey qubits are the ones that represent terms in a logical Hamiltonian, whereas the other parity qubits are auxiliary and their local fields are tuned to $0$.} 
    \label{fig: built LHZ}
\end{figure}

We will now briefly explain how to construct a graphical representation of the Hamiltonian $\tilde{H}$ to easily find such loops. We start by inspecting the indices of the physical spins, $\tilde{\sigma}_z^{(i,j)}$. The most general case is when all interactions between spins are present in $H$, or, in other words, if its associated graph representation has all-to-all connectivity. In \cref{fig: built LHZ}, a graphical representation of the description below is shown. We start by placing the physical spins that represent the logical local fields, $\tilde{\sigma}_z^{(i)}$, in a row. We then draw diagonal perpendicular lines (where each line represents one logical spin) in a grid, so that each intersection between lines is an interaction term in the Hamiltonian. Hence, each physical spin $\tilde{\sigma}_z^{(i, j)}$ is placed at the intersection of the lines that start from $\tilde{\sigma}_z^{(i)}$ and $\tilde{\sigma}_z^{(j)}$. By filling all the intersections this way, we obtain a triangular system of spins that represents a logical Ising Hamiltonian with all-to-all connectivity. We now exploit the squares and triangles between the intersecting lines to identify the physical spins in each plaquette to construct our parity constraints, $H_P$. We place a triangle between the physical spins subject to a $3$-body interaction term and a square between the qubits subject to a $4$-body interaction term. In this way, we obtained a quantum system that represents the Hamiltonian $\tilde{H}$. 

To implement $H_l$ we now only need to program the local fields of the appropriate physical spins. Thus, the local fields of the spins $\tilde{\sigma}_z^{(i)}$ are set to $h_i$ and the local fields of the spins $\tilde{\sigma}_z^{(i,j)}$ are set to $J_{i,j}$. 

\subsection{Generalized parity embedding}\label{subsec: parity map}
We can generalize the LHZ triangle and find parity Hamiltonians for arbitrary Ising Hamiltonians with high-order interactions and with a smaller number of physical spins than shown in the previous section~\cite{ender2023parity}. By considering \cref{eq: parity constraint condition}, we can define a condition to preserve parity by inspecting the indices of the terms directly, instead of considering the loops in the graph. Given an Ising 
Hamiltonian with high-order interactions:
\begin{multline}\label{eq: ising ho}
    H = \sum_i h_i\sigma_z^{i} + \sum_{i< j} J_{i,j} \sigma_z^i\sigma_z^j + \cdots +\\+ \sum_{i_1,\ldots,i_n} J_{i_1,\ldots, i_n}\sigma_z^{i_1}\cdots\sigma_z^{i_n},
\end{multline}
we can apply the same map that associates each term of the Hamiltonian $H$ to a physical spin:
\begin{align*}
    \sigma_z^{i}&\mapsto\tilde{\sigma}_z^{(i)}\\    \sigma_z^i\sigma_z^j&\mapsto\tilde{\sigma}_z^{(i,j)}\\    &\cdots\\    \sigma_z^{i_0}\cdots\sigma_z^{i_n}&\mapsto\tilde{\sigma}_z^{(i_0,\ldots, i_n)}.
\end{align*}
Notice that \cref{eq: parity constraint condition} always holds if the number of times an index appears is even. Therefore, we can use the same argument for an arbitrary number of indices. Hence, for instance, if the terms $\sigma_z^i,\; \sigma_z^j\sigma_z^k,\; \sigma_z^i\sigma_z^p$ and $\sigma_z^j\sigma_z^k\sigma_z^p$, appear in $H$, we have that
\begin{equation*}
\sigma_z^i\sigma_z^j\sigma_z^k\sigma_z^i\sigma_z^p\sigma_z^j\sigma_z^k\sigma_z^p=1.
\end{equation*}
Therefore, to ensure that the parity is preserved we must fulfill
\begin{equation}\label{eq: parity condition ho}
    \tilde{\sigma}_z^{(i)}\tilde{\sigma}_z^{(j, k)}\tilde{\sigma}_z^{(i, p)}\tilde{\sigma}_z^{(j,k,p)} = 1
\end{equation}
for all $i, j, k$ and $p$. Extending to higher dimensions, we move away from the graphical representation and rewrite \cref{eq: parity condition ho} as follows: let $\ket{\tilde{\psi}}$ be a wavefunction of the Hilbert space defined by the terms $\tilde{\sigma}_z^{(i_0,\ldots, i_n)}$. It is straightforward to see that if $\ket{\tilde{\psi}}$ satisfy \cref{eq: parity condition ho} then 
\begin{equation}\label{eq: parity autospace}
    \tilde{\sigma}_z^{(i)}\tilde{\sigma}_z^{(j, k)}\tilde{\sigma}_z^{(i, p)}\tilde{\sigma}_z^{(j,k,p)}\ket{\tilde{\psi}}=\ket{\tilde{\psi}}.
\end{equation}
Thus, a solution is valid if and only if this condition is fulfilled. Therefore, to reduce the degrees of freedom in the system we can use $4$- and $3$-body interactions as before.

However, given this generalization, finding the right set of indices to constrain the system (such that each physical spin is included in at least one of the $K-N$ constraints) is not a trivial problem. The different classical techniques needed to find a suitable arrangement of the qubits and plaquettes are an active area of research~\cite{ter2023constructive, ter2023flexible}.

\subsection{Parity compiled Hamiltonian with at most quadratic order interactions}\label{subsec: QUBO LHZ}
Direct implementations of a parity-compiled Hamiltonian necessarily require $4$- and $3$-body interaction terms. These kinds of interactions cannot be implemented as native operations in current quantum annealers (although new such dedicated hardware is being built~\cite{NEC}). However, a solution to express \cref{eq: parity constraint ham} by using at most quadratic interaction terms has been proposed in~\cite{rocchetto2016stabilizers}, which we briefly review.

We start by observing the following: consider the Hamiltonian describing the square parity plaquette $l$
\begin{equation*}
-\tilde{\sigma}_z^{(n,l)}\tilde{\sigma}_z^{(e,l)}\tilde{\sigma}_z^{(w,l)}\tilde{\sigma}_z^{(s,l)}.
\end{equation*}
We see that in all ground states the number of $\ket{\uparrow}$ states is even--- we call this the even parity constraint. Likewise, if we were to have a Hamiltonian consisting of the term
\begin{equation*}
\tilde{\sigma}_z^{(n,l)}\tilde{\sigma}_z^{(e,l)}\tilde{\sigma}_z^{(w,l)}\tilde{\sigma}_z^{(s,l)},
\end{equation*}
we would observe that in all ground states the number of $\ket{\uparrow}$ states is odd--- we call this the odd parity constraint.
Moreover, the degeneracy of the ground state of these two Hamiltonians is the same. Thus, there is a one-to-one correspondence between the ground states of the two Hamiltonians, which can be achieved by flipping the value of one spin. This method of flipping spins conserves the properties of the ground state distribution of a given Ising Hamiltonian~\cite{DWave2009doc}. Therefore, we can rewrite $H_P$ as\footnote{This same argument can be applied to triangular plaquettes.}
\begin{equation}\label{eq: odd parity ham}
    H_P' = -H_P = \sum_{l=0}^{K-N-1} \tilde{\sigma'}_z^{l, n}\tilde{\sigma'}_z^{l,e} \tilde{\sigma'}_z^{l, w} \left(\tilde{\sigma'}_z^{l,s}\right),
\end{equation}
where one of the qubits $\tilde{\sigma'}_z^{l,i}$ is flipped, hence $\tilde{\sigma'}_z^{l,i} = -\tilde{\sigma}_z^{l,i}$ for $i\in\{n,e,w,s\}$ and $\tilde{\sigma'}_z^{l,j} = \tilde{\sigma}_z^{l,j}$ for $j\ne i$.
Under this change of sign, we stress that the two formulations are equivalent. Consider the plaquette $\bar{l}$ and the parity qubit with flipped local field $\tilde{\sigma'}_z^{l, w}$, then the following holds \footnote{We keep using the round bracket notation when we want to consider $4$- and $3$- body terms simultaneously.}
\begin{multline*}
    \tilde{\sigma'}_z^{\bar{l}, n}\tilde{\sigma'}_z^{\bar{l},e} \tilde{\sigma'}_z^{\bar{l}, w} \left(\tilde{\sigma'}_z^{\bar{l},s}\right) = \\ = \tilde{\sigma}_z^{\bar{l}, n}\tilde{\sigma}_z^{\bar{l},e} \left[-\tilde{\sigma}_z^{\bar{l}, w}\right] \left(\tilde{\sigma}_z^{\bar{l},s}\right) = -\tilde{\sigma}_z^{\bar{l}, n}\tilde{\sigma}_z^{\bar{l},e} \tilde{\sigma}_z^{\bar{l}, w} \left(\tilde{\sigma}_z^{\bar{l},s}\right).
\end{multline*}

\begin{figure}
    \centering
    \subfloat[Parity constraints]{\includegraphics[scale=0.15]{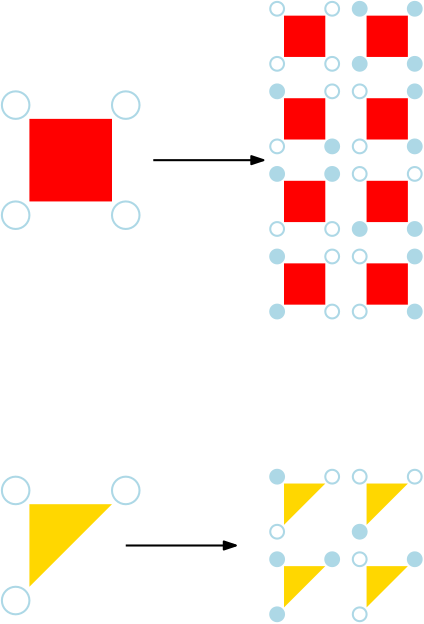}}\hfill\subfloat[Odd parity constraints]{\includegraphics[scale=0.125]{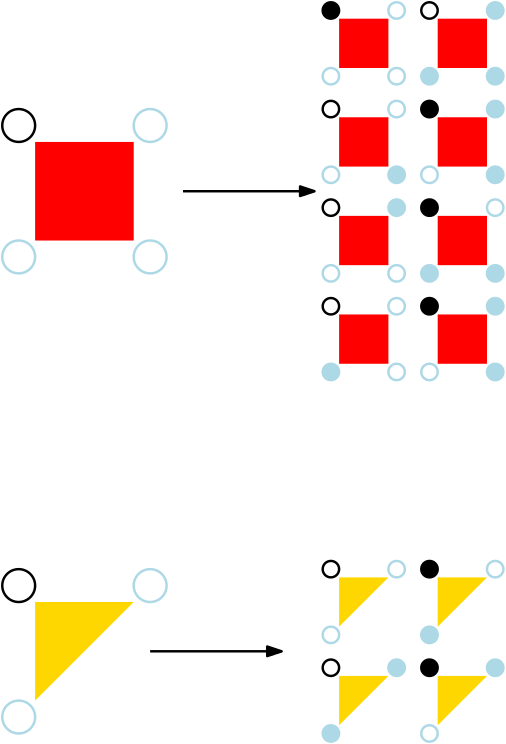}}\hfill\subfloat[quadratic odd parity constraints]{\includegraphics[scale=0.11]{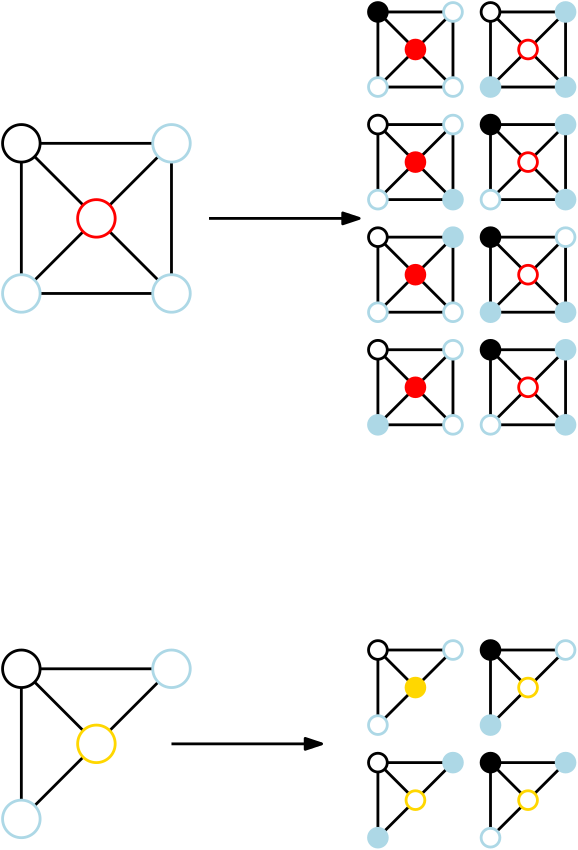}}
    
    \caption{Visualization of different implementations of parity constraints. The red squares and the yellow triangles represent the $4$- (\emph{top}) and $3$-body (\emph{bottom}) interactions of: \textbf{a)} the parity constraints Hamiltonian as in \cref{eq: parity constraint ham} and \textbf{b)} the odd parity constraints Hamiltonian as defined in \cref{eq: odd parity ham}. In \textbf{c)}, the yellow and red dots represent instead the auxiliary qubits used to write the Ising formalization of the $4$- and $3$-body interactions as in \cref{eq: quadratic odd parity ham}. In this case, $2$-body interactions are represented by edges between nodes. In each diagram, all the possible ground states obtainable from the plaquettes are shown. The filled dots represent spins in the state $\ket{\uparrow}$, while the white dots represent spin in the state $\ket{\downarrow}$. The black dots in the odd parity constraints represent the flipped local field qubits. Each of the possible ground states shown on the right-hand side of each diagram is displayed such that equivalent ground states written using a different parity constraints formalism each have the same location in their corresponding diagram. For instance, the first square on the top left corner of each series of ground states is the same state but written using a different parity constraints formalism.}
    \label{fig: odd and qubo contraints}
\end{figure}

In \cref{fig: odd and qubo contraints}, we give a visualization of the even and odd parity constraints both for squares (\textit{top}) and triangles (\textit{bottom}). We can see that the ground states of the even parity plaquettes (\textit{left}) and the odd parity plaquettes (\textit{center}) are equivalent.

Now, from the odd parity constraints formulation we can define a new Hamiltonian with at most quadratic interaction terms. Since we know that we cannot directly implement higher-order interactions with quadratic interactions, we have to introduce auxiliary qubits. In this case, only one qubit is required to replace the $4$- and $3$-body interactions. Let $\tilde{\sigma'}_z^{l, a}$ be the auxiliary qubit of the plaquette $l$, we can rewrite \cref{eq: odd parity ham} as
\begin{multline}\label{eq: quadratic odd parity ham}
    H_P = \sum_{l:\;\textrm{$l$ is $\square$}}\left(2\tilde{\sigma'}_z^{l, a} + \tilde{\sigma'}_z^{l,n}+\tilde{\sigma'}_z^{l,e}+\tilde{\sigma'}_z^{l,w}+\tilde{\sigma'}_z^{l,s}\right)^{2}+\\+\sum_{l:\;\textrm{$l$ is $\bigtriangleup$}}\left(1+2\tilde{\sigma'}_z^{l,a} + \tilde{\sigma'}_z^{l,n}+\tilde{\sigma'}_z^{l,e}+\tilde{\sigma'}_z^{l,w}\right)^2.
\end{multline}
Notice that we keep the notation $\tilde{\sigma'}_z$ because in each plaquette at least one local field is flipped to implement the odd parity constraints. 
As with the even and odd constraints, in \cref{fig: odd and qubo contraints} we can see the equivalence between the ground states of the parity plaquettes defined by using quadratic order interactions.

It must be noted that finding a set of parity qubits (which implement the odd constraints) to flip for general parity compiled problems is a hard problem\footnote{This can be reduced to the minimum vertex cover problem.}. However, this problem becomes trivial if the parity compilation consists only of square plaquettes, due to the regular grid structure. In this specific case, we flip every second qubit in the grid lattice to find a solution to this problem. 

\begin{figure}
    \centering
    \includegraphics[scale=0.2]{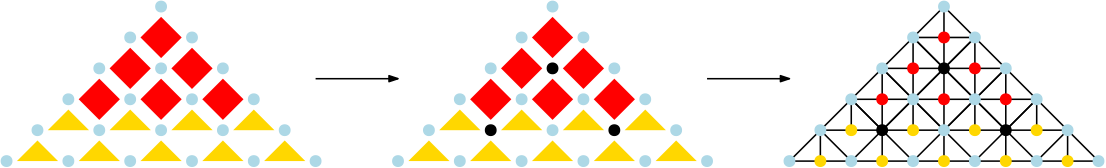}
    \caption{Three different visual representations of parity Hamiltonians. We can start from a parity Hamiltonian with even parity constraints as defined in \cref{eq: parity constraint ham}. Then, by flipping every second parity qubit we can implement the square plaquettes and some triangular plaquettes as odd parity constraints like in \cref{eq: odd parity ham}. Eventually, we rewrite the Hamiltonians to use only $2$-body constraints and we obtain a version of the parity compiled Hamiltonian with only $2$-body interactions.}
    \label{fig: qubo LHZ}
\end{figure}

\begin{figure}
    \centering
    \includegraphics[scale=0.2]{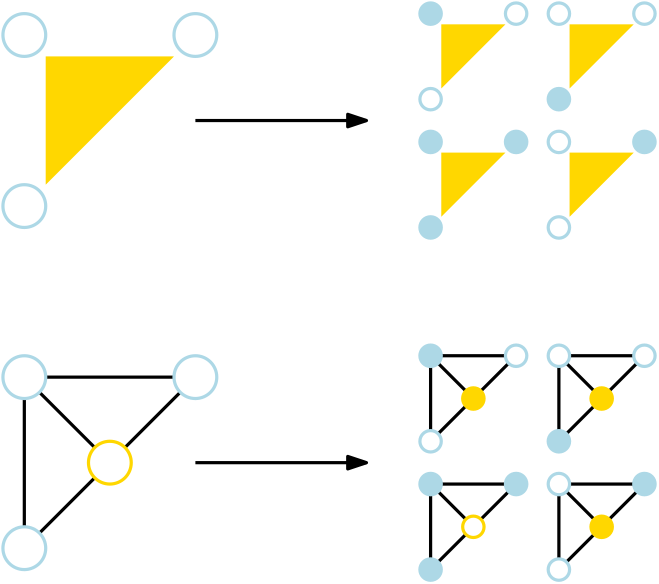}
    \caption{We can find an Ising Hamiltonian with only two-qubit interactions for the even triangular parity plaquette. On the right hand side of this figure, the ground states of the $3$-body and $2$-body plaquette are shown. Equivalent ground states of each of the parity configurations are placed in corresponding locations. The light-blue dots represent the parity qubits involved in the plaquette and the yellow dot is the auxiliary qubit used to implement the same parity plaquette with only two-qubit interactions.}
    \label{fig: qubo triangle}
\end{figure}

To complete the implementation of the parity constraints, we note that the even triangle constraint \cref{eq: parity constraint ham} can also be implemented with quadratic interaction terms and only one auxiliary spin, as follows: 
\begin{equation*}
    \left(1-2\tilde{\sigma}_z^{l, a} - \tilde{\sigma}_z^{l,n}-\tilde{\sigma}_z^{l,e}-\tilde{\sigma}_z^{l,w}\right)^2.
\end{equation*}
Note that the notation $\tilde{\sigma}_z$ is used in this case because no parity qubit in the plaquette is flipped.
Thus, in the case that the parity compiled problem contains triangles, as in \cref{fig: qubo LHZ}, we can use the even or odd triangle parity constraints interchangeably. This is advantageous because both the odd and even triangle parity constraints require the addition of only one auxiliary spin. The same is not true for the even square parity constraints, since its formulation with only quadratic terms requires two auxiliary spins, which increases the overall spin count. 
On the top left-hand side of \cref{fig: qubo triangle} we see the even $3$-body triangle parity constraint. Below we see its transformation to a constraint with only 2-body interactions. To the right of both of the diagrams, we see the corresponding ground states of both of the constraints.
Henceforth, we refer to the parity compiled Hamiltonian with $4$- and $3$-body interactions as a multi-body parity compilation, whereas we refer to its version with only $2$-body interactions as a $2$-body parity compilation.

\section{PQC on D-Wave annealers}\label{sec: parity on dwave}
In this section, we show how to extend the parity mapping described in \cref{subsec: QUBO LHZ} to a quantum annealer with fixed topology with at most quadratic interaction terms. To implement a parity compiled Hamiltonian on fixed topology quantum annealers, specifically D-Wave Pegasus topology, we must embed the $2$-body parity compiled problem into the Pegasus hardware topology. Even though one can find such embedding by using common minor embedding techniques, in this work we present two different scalable embeddings that do not require any additional compilation in the quantum annealers topology. Thus, those scalable frameworks allow the implementation of any parity compiled problem without finding a minor embedding. Moreover, Ising Hamiltonians consisting of at most $2$-body interactions can be implemented as $2$-body LHZ triangles without the need to use additional computational resources, neither a parity compilation nor a minor embedding technique.

At the time of writing, D-Wave's annealer topology is based on the Pegasus graphs~\cite{boothby2020next}. Pegasus graphs $P_n$ are a family of modular graphs fulfilling the property $P_{n-1}\subset P_n$. Therefore, if we embed a repetitive structure in the smallest Pegasus graph, we obtain an embedding for $P_n$ as well, for all $n$.

Considering the geometry of the $2$-body square parity constraint, i.e. the unweighted undirected graph, we can see that this graph is isomorphic to a complete graph with $5$ nodes, $K_5$. Analogously, the $2$-body triangle parity constraints are isomorphic to the complete graph with $4$ nodes, $K_4$. Therefore, since $K_4$ is a subgraph of $K_5$, the $2$-body triangular plaquette is a subgraph of the $2$-body square plaquette. Thus, if we consider a modular graph built using the $2$-body square plaquette as a minimal component, we obtain a graph where we can implement any $2$-body parity compiled problem. Therefore, to find an embedding for parity compiled problems that covers the whole hardware graph, we have to find a modular and repetitive subgraph of the Pegasus graph where the $2$-body square plaquette can be embedded. 

\begin{figure}
    \centering
    \includegraphics[scale=0.2]{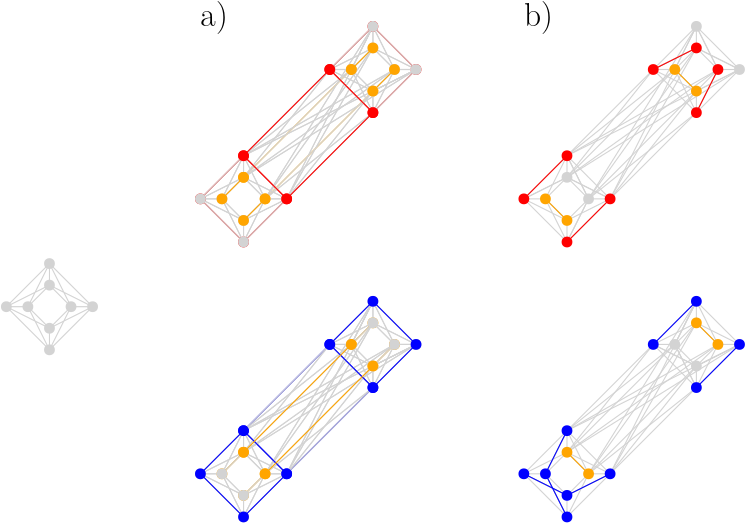}
    \caption{Considering the orthogonal projection on a plane of the Pegasus graph, we can identify unit cells of $8$ qubits called diamonds. On the left of the picture, such a structure is highlighted. In the quantum annealers graph, rows of qubits can be identified and numbered. If we look at the connection between adjacent diamonds in the same row, we can define the structure of our embeddings.
    In this figure, the red (blue) qubit chains are the physical parity qubits on the rows numbered with even (odd) numbering in the quantum annealers. Whereas, the orange qubit chains implement the auxiliary qubits needed for the quadratic parity constraints.
    \textbf{a)} We can use loops in the diamonds and connections of the diamonds to implement the parity qubits. The embedding using these structures is called original. Or, \textbf{b)} We can decide to implement one square plaquette spread on two different diamonds. The embedding built in this way is called dense.}
    \label{fig: diamonds}
\end{figure}

\begin{figure}
    \centering
    \includegraphics[scale=0.3]{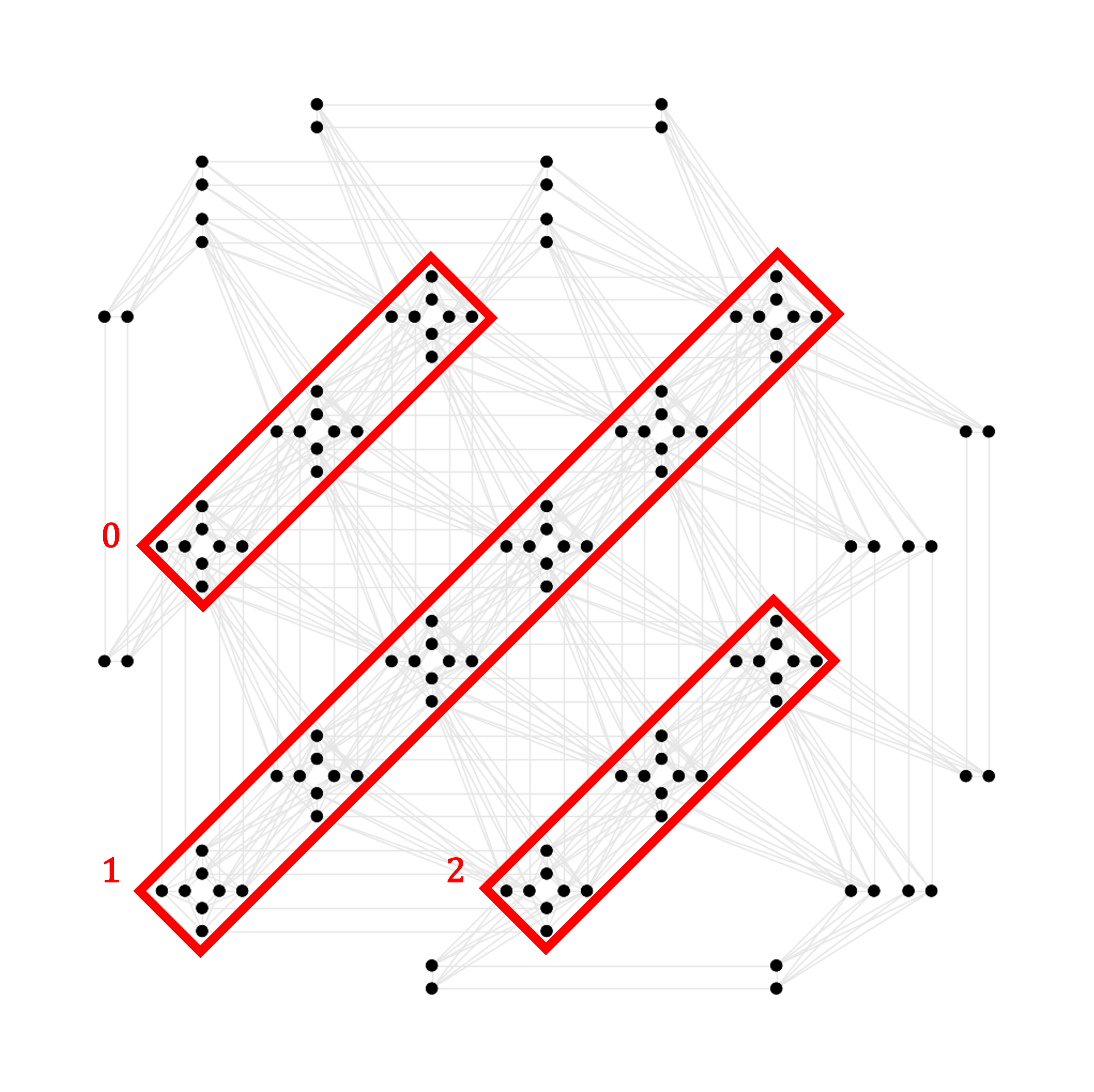}
    \caption{Pegasus graph $P_3$ and its orthogonal projection into a plane. We highlighted the rows of repetitive diamonds as explained in the main text. We can number the row of $P_n$ starting from the one closest to the top left corner ($0$) to the one closest to the bottom right corner ($2(n-2)$). We define the upper (lower) side of the picture to be the top left (bottom right) corner. The left (right) side of the picture is, therefore, the bottom left (top right) corner of the picture.}
    \label{fig: rows}
\end{figure}

We show how to find such a subgraph by using two adjacent square plaquettes as the minimal component of our repetitive structured graph. If we consider the Pegasus graph in its orthogonal projection, we can use groups of $8$ qubits called diamonds that are the unit cells of the repetitive graph, see \cref{fig: diamonds} (\textit{left}). 
These diamonds are arranged in rows that can be numbered (we start numbering from $0$), see \cref{fig: rows}. We characterize the chains of spins in a diamond based on the number associated with that row. Thus, we can create an alternation of physical spin chains that have the desired connectivity to be able to embed the $2$-body square plaquettes.

By considering two consecutive diamonds in the same row, we are able to define an embedding for the $2$-body parity plaquettes, by identifying loops of $4$ physical spins which define a parity qubit, as shown in \cref{fig: diamonds} \textbf{a)}. 
%We call this embedding ``original''.
%
By exploiting these loops and the connections between them, we can define an embedding for the $2$-body parity plaquettes--- we call this embedding ``original''. 

\begin{figure*}
    \centering
    \subfloat{\includegraphics[]{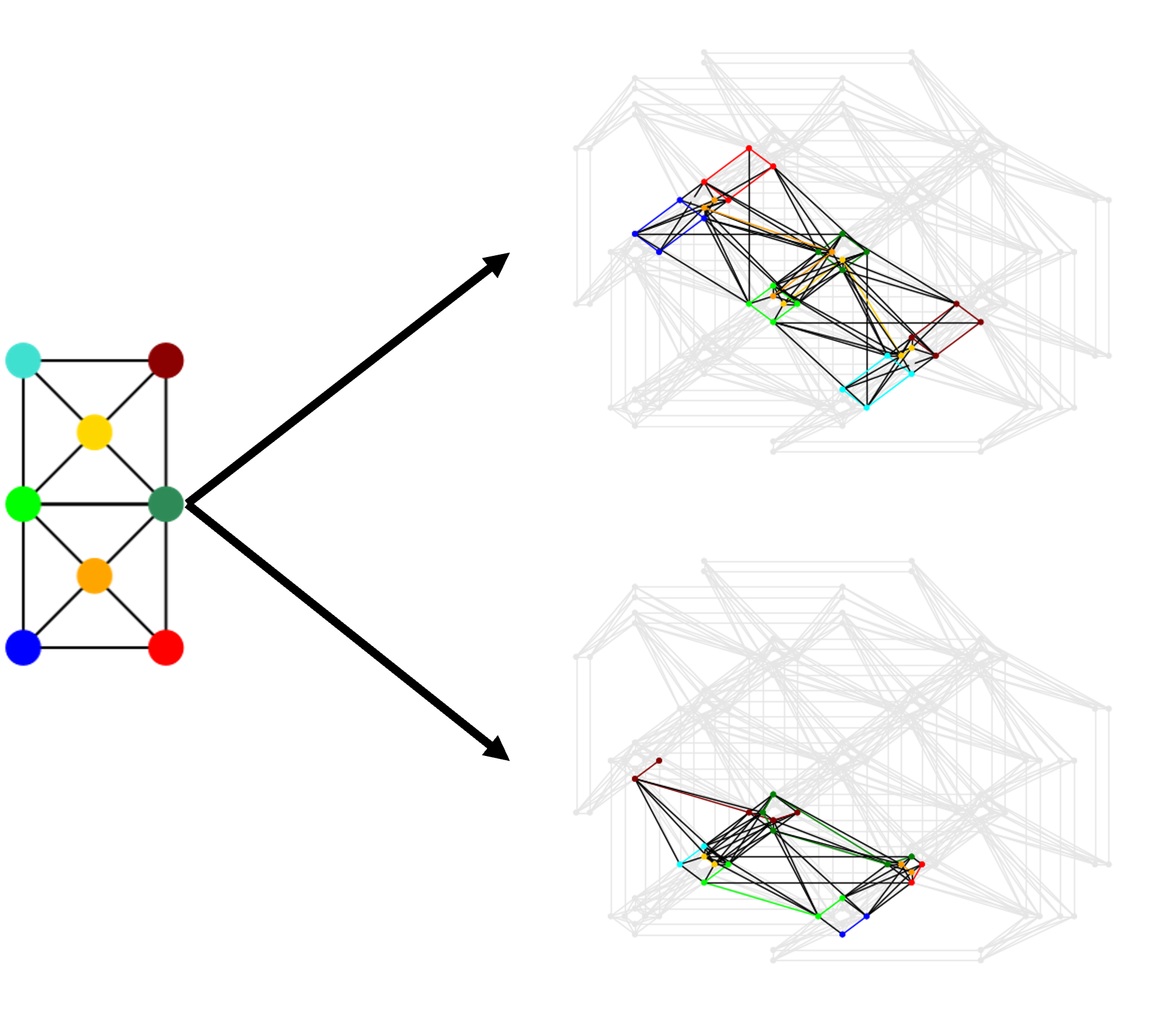}}

    \subfloat{\includegraphics[scale=0.45]{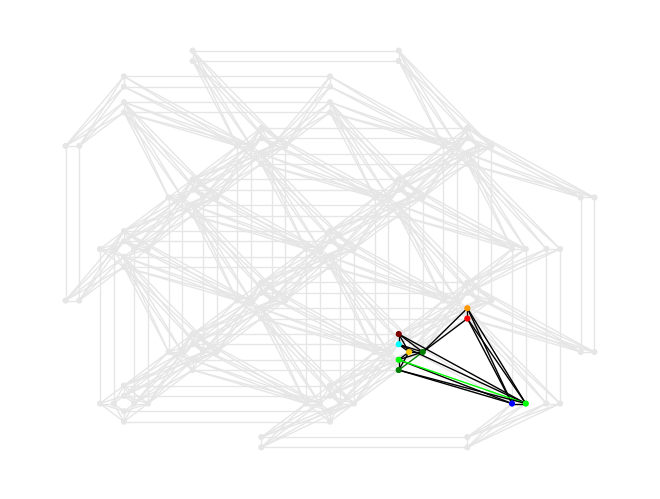}}\hfill\subfloat{\includegraphics[scale=0.45]{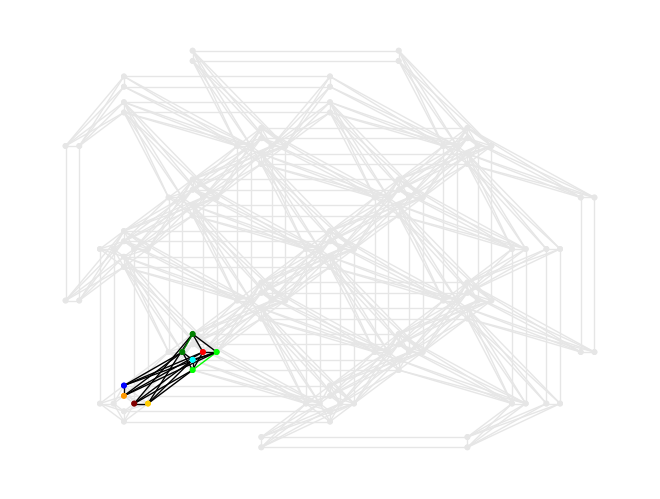}}
    \caption{The top diagram shows two different ways to embed the quadratic parity constraints into the D-Wave's quantum annealer architecture. These implementations are obtained by considering the embeddings of the qubits as described in \cref{fig: diamonds}. In the picture, the correspondence between each qubit of two quadratic square plaquettes and the chain of physical spins in the quantum annealers is shown by means of colors. The upward arrow indicates the original embedding, whereas the downward arrow indicates the dense embedding.
    In the bottom diagram, two alternative embeddings of the two square plaquettes are shown. These are outcomes from the minor embedding function. Even though the number of spins used in the implementation is smaller in the found embeddings, they are not scalable and cannot be used to cover the whole hardware graph. We can see that the geometry of these graphs is not compatible with the symmetries of the Pegasus graph and they cannot be used to create a modular and repetitive structure in the hardware graph. Notice that the shown embeddings are chosen arbitrarily since the minor embedding function is not deterministic, hence it has multiple potential outputs. Thus, finding such a modular and repetitive structure to cover the whole hardware graph by using minor embedding techniques would be hard since the heuristic is not thought to solve the minor problem by considering such properties of the graph.}
    \label{fig: embeddings}
\end{figure*}

In the even rows, we can implement a parity qubit by considering the $4$-spin loop that can be obtained by connecting the $2$ outer spins of two adjacent diamonds facing each other. Whereas, in the odd rows, we can use the $4$ external spins of the diamond as a loop. On the other hand, to implement the auxiliary qubits, we can consider a chain of spins which consists of the $4$ internal spins of the diamonds.
In an even-numbered row, we consider the two internal spins on the 
upper\footnote{The upper (lower) side of a diamond is the $4$ spins of the diamond in the row $i$ that faced row $i+1$ ($i-1$). Likewise, the upper (lower) plaquette of the $i$-th row is the plaquette defined on the rows $i$ and $i+1$ ($i-1$). } 
(lower) side of the diamond as part of the group of spins that form the chain for the auxiliary qubit of the upper (lower) square plaquette. 
In an odd-numbered row, the upper (lower) auxiliary qubit is implemented by taking the top (bottom) left internal spin of the left diamond and the top (bottom) right internal spin of the right diamond. 
Hence, a $2$-body square plaquette is implemented in two different rows: two parity qubits are contained as $4$-spin loops in an even-numbered row and the other two parity qubits are contained as $4$-spin loops in an odd-numbered row; the auxiliary qubit is implemented as a $4$-spin chain that involves two spins of the even-numbered row and two spins from the odd-numbered row. The `original' embedding of two adjacent square plaquettes is indicated by the upwards arrow in \cref{fig: embeddings}.

In contrast to the original embedding, instead of considering chains consisting of loops of spins, we can derive a different embedding, by first implementing a $2$-body square plaquette in two adjacent diamonds on the same row. Since this embedding uses fewer physical spins, we call this embedding ``dense''.  If we consider two internal spins on the left\footnote{Left (right) side of the diamond is defined to be the one which faces to the bottom left (top right) corner of the picture.} (right) side of a diamond, we can notice that they are connected to all of the external spins of the same diamond to which they belong and to the $4$ external spins of the adjacent left (right) diamond. Thus, by using these two spins as a chain we can implement the auxiliary qubit. By following this argument, we can use the other spins in the diamonds to implement the parity qubits as chains of $2$ and $3$ spins within the diamond. In an even-numbered row, to implement the auxiliary qubit we use the left internal spins. Then, we can alternate a diamond where two parity qubits are defined as $2$-spin chains that connect two right external spins to the two left external spins and a diamond where the parity qubits are defined as $2$-spin chains that connect the upper (lower) left external spins to the upper (lower) right internal spins, which are available because they are not used to implement the auxiliary qubit. Instead, in an odd-numbered row, to implement the auxiliary qubit we use the right internal spins. Then, we alternate a diamond where $3$-spin chains connect the two antipodal external spins in a diamond passing through a left internal spin, which is available since it is not used to implement the auxiliary qubit and a diamond where the two left external spins are connected with a $2$-spin chain with the left external ones. Eventually, we can identify the parity qubit represented by the chain containing the bottom left physical spin of a diamond in a (odd-) even-numbered row with the parity qubit represented by the chain containing the top left spin of a diamond in a (even-) odd-numbered row. With this identification, we create longer chains that generate a grid of parity qubits. Therefore, every diamond contains exactly two parity qubits, where one is shared between adjacent rows. The embedding of two adjacent plaquettes is indicated by the downwards arrow in \cref{fig: embeddings}.

Furthermore, the new proposed embeddings create a new graph topology for the quantum annealing hardware. Considering the qubits implemented as chains of spins, we can derive two different hardware graphs, as we can see in \cref{fig: new topo}. These topologies could be used to either implement LHZ triangles or parity compiled problems that fit the new hardware graph.
\newline

During the writing of this manuscript, we become aware of similar techniques to implement LHZ triangles in D-Wave quantum annealers~\cite{Miyazaki_2024}.

\section{Experiments and results}\label{sec: results}
In this section, we present the outcomes of the experiments and benchmarks. 
We study how the performance of the quantum annealers influences the probability of measuring the ground state in basic parity compiled problems. Moreover, we deeply analyze the distribution of the ground states and compare it to the theoretical distribution of valid configurations. Furthermore, we simulate different implementations of several instances of the multi-car paintshop problem encoded as Ising Hamiltonians and we study their minimum spectral gaps. 
Eventually, we compare locations of the minimum spectral gaps of simulated and experimental implementations of the multi-car paint shop problem. The experimental problems are implemented using the developed embeddings and solved using QA.

The results are obtained from the quantum annealing hardware ``Advantage system 4.1" and numerical simulations on CPUs.

\subsection{Test-bed: the multi-car paint shop problem}\label{sec: psp}
To test the idea presented in \cref{sec: parity on dwave} we choose a combinatorial optimization problem that both has a natural implementation as a frustrated Ising chain and is of practical relevance to the automotive industry: the multi-car paint shop problem~\cite{yarkoni2021multi}. We briefly review this problem, explain its real-world relevance, and examine its computational complexity.

In the process of manufacturing cars, a crucial step is the painting of the car body. After the car body is assembled, it is arranged on a assembly line with a fixed order of the bodies, based on customer orders. Due to production constraints, the order of the car bodies cannot be permuted once on the line. The bodies enter the paint shop sequentially and go through two different treatments. They are first painted using a coating known as the filler, an initial protective coat of paint that is either dark or bright (black or white). After this, the car bodies are painted with the final body color. Since the customer orders are received randomly, the order of the colors of the car bodies is also random. Whenever a change of color occurs between two consecutive bodies, the nozzle of the paint guns must be cleaned, resulting in a waste of paint and cleaning materials, which incurs a monetary cost. Thus, it is of practical interest to minimize the number of color switches, which can be done by virtually swapping the colors assigned to identical car bodies (thus not permuting the order of bodies physically)~\cite{streifbps,yarkoni2021multi}. It has been shown that for this definition of the paint shop problem, minimizing the number of switches is NP-hard and that the problem belongs to the APX-hard class~\cite{bonsma2006complexity}. This makes it of particular interest for investigation with quantum algorithms.

In this work, we focus on the optimization of the filler in groups of cars; meaning, we have only two colors and groups of cars of different types. To implement the problem as an Ising Hamiltonian, we can look at groups of cars of the same model but that are assigned to different fillers. We can see that switching the filler between those cars does not change the order of the bodies in the chain but only the order of the colors. We can write this problem as a frustrated Ising chain where each spin $s_i$ represents a car $i$, the state $s_i = 1~(s_i = 0)$ represents the car being painted black (white). Thus, we want to minimize the number of anti-ferromagnetic pairs in the chain, i.e.
\begin{equation}\label{eq: of Ham}
    H_\textrm{of} = -\frac{1}{C-1}\sum_{i=0}^{C-2}s_is_{i+1},     
\end{equation}
where $C$ is the number of cars in the line of bodies. Then, we can identify the sets of cars $G_j$ of the same model, where $k_j\in\mathbb{N}$ cars must be painted in black. By adding this condition to~\cref{eq: of Ham}, we obtain the Ising Hamiltonian
\begin{equation}\label{eq: psp}
    H = H_\textrm{of} + \lambda\sum_j\left[(|G_j| - 2k_j)\sum_{i\in G_j}s_i + \frac{1}{2}\sum_{i,j\in G_j}s_is_j\right],
\end{equation}
where $|G_j|$ is the cardinality of the set $G_j$, i.e. the number of model $j$ cars. We say a problem to be trivial if $|G_j|=k_j$ for some $j$ set of cars.

In this work, we solve all possible non-trivial Hamiltonians that implement the multi-car paint shop problem with $2$ to $5$ cars. 

\begin{figure*}[!ht]
    \centering
    \includegraphics[scale=0.55]{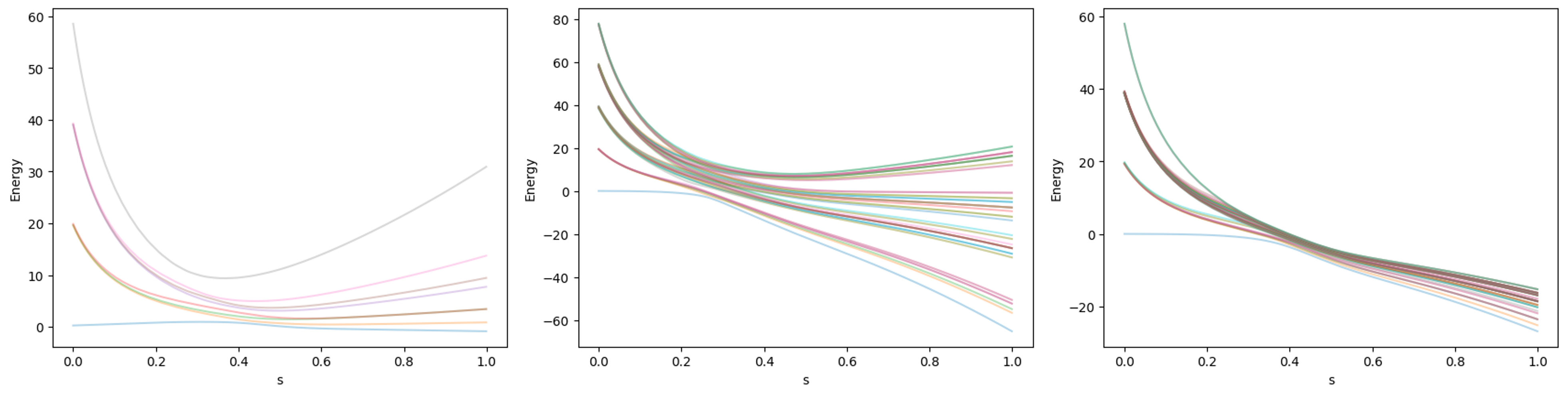}
    \caption{Simulated dynamics of the instantaneous eigenvalues of the time-dependent Hamiltonian that represents the QA evolution that solves a paint shop problem as described in \cref{sec: psp} with $3$ cars, $|G_0|=3$ and $k_0=1$. The Hamiltonian of the problem is implemented in three different ways: (\textit{on the left}) the problem Hamiltonian is a frustrated Ising chain; (\textit{in the center}) the problem Hamiltonian is a multi-body LHZ triangle; and, (\textit{on the right}) the problem Hamiltonian is a $2$-body LHZ triangle. Notice that due to the redundant encoding, the LHZ implementations present more instantaneous eigenvalues. For clarity, in the pictures presenting the LHZ eigenvalues dynamics only the first $50$ lowest eigenvalues are plotted. In this work, we consider only the minimum distance between the lowest and the second-lowest eigenvalues. Therefore, no further considerations are made on the full dynamics of the eigenvalues.}
    \label{fig: dyn eig}
\end{figure*}

\subsection{Numerical Results}\label{sec: simulation}
As presented in \cref{sec: QA}, the performance of QA is linked to the minimum spectral gap of the matrix that represents the evolution of the quantum system. Therefore, to have an idea of the performance of the algorithm and how this could be influenced by the parity transformation, we simulate the dynamics of the instantaneous eigenvalues of the problem as described in \cref{sec: psp} in three different implementations: frustrated Ising chain, multi-body LHZ triangle, and $2$-body LHZ triangle.

Since there may be multiple optimal solutions to a multi-car paint shop problem, the ground state of the frustrated Ising Hamiltonian is degenerate. Thus, if the initial Hamiltonian does not match the degeneracy of the problem Hamiltonian, the spectral gap of the time-dependent Hamiltonian that describes QA vanishes. Therefore, since we can only implement the initial Hamiltonian with a non-degenerate ground state, we have to slightly change the paint shop problem Hamiltonian to make its ground state non-degenerate. We can achieve such a setting by adding a ``small'' bias on the local field of the last spin in the frustrated Ising chain.

Furthermore, to have a fair comparison between the different implementations we set the penalty coefficient $\lambda$ of \cref{eq: psp} to be $1$ and we compute the optimal coefficient of parity constraint Hamiltonian such that the first two lowest eigenvalues are separated from the rest of the eigenstates. To compute such coefficients we use~\cite{lanthaler2021minimal}.

To compute the instantaneous eigenstate trajectories, we store the Hamiltonian representing the evolution as a matrix, we diagonalize it at different time steps and store the eigenvalues. Thus, we can plot the dynamics of the individual eigenvalues and we can compute the minimum spectral gap point. In \cref{fig: dyn eig}, we can see an example of the dynamics of the instantaneous eigenvalues for a particular instance when the instance is implemented in three different ways. First as a frustrated Ising chain, second as a multi-body LHZ triangle and third as a 2-body LHZ triangle. Henceforth we focus only on the study of the minimum spectral gap point and its value, but we do not consider the whole dynamics of the instantaneous eigenstate since it is out of the scope of this work.

\begin{figure}
    \centering
    \includegraphics[scale=0.45]{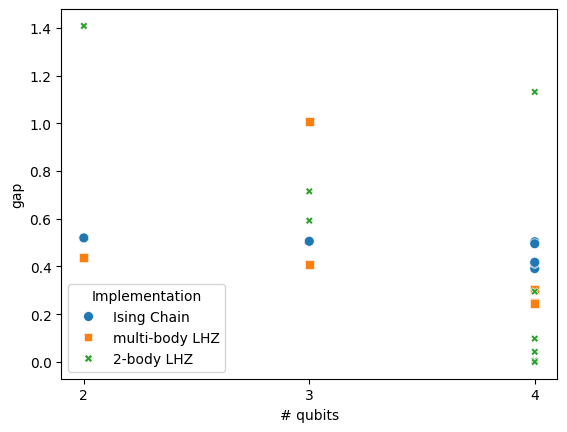}
    \caption{Values of the minimum spectral gaps for different instances of multi-car paint shop problem as described in \cref{sec: psp} and different implementations. We can see that by implementing the problem as a frustrated Ising chain (\textit{blue circles}) the values of the minimum spectral gaps for different instances do not differ much, whereas if we implement the problem as an LHZ triangle (\textit{orange squares} for multi-body LHZ and \textit{green crosses} for $2$-body LHZ) the value of the minimum spectral gap can vary significantly depending on the instance. Thus, the redundant encoding of the LHZ does not necessarily lead to a shrinking of the minimum spectral gap value. Furthermore, we can see that some instances of the problem result in a larger spectral gap when implemented as LHZ Hamiltonian.}
    \label{fig: min spectral gaps}
\end{figure}

In \cref{fig: min spectral gaps}, we can compare the different minimum spectral gap values against problem size (number of qubits) for the three different problem implementations. Notice that although for the frustrated Ising chain implementation the values of the gaps are similar, the same cannot be said for the other two LHZ implementations. We can conclude that, in general, the spectral gap does not necessarily shrink if we implement the paint shop problem as an LHZ triangle. Especially for some specific instances, we can see that the LHZ implementations can get better results. Therefore, writing a combinatorial optimization problem as a parity compiled Hamiltonian does not decrease the performance of QA a priori since the value of the minimum spectral gap is not necessarily decreasing. Furthermore, different spectral gap values could be found by considering different compilation techniques employing the parity mapping. Therefore future research should focus on finding the best geometry of the parity compilation per class of optimization problems or instances. 

\subsection{Experimental Results}\label{subsec: exp}
\begin{figure*}
    \centering
    \subfloat[New topology of a $P_4$ defined by the original embedding]{\includegraphics[scale=0.35]{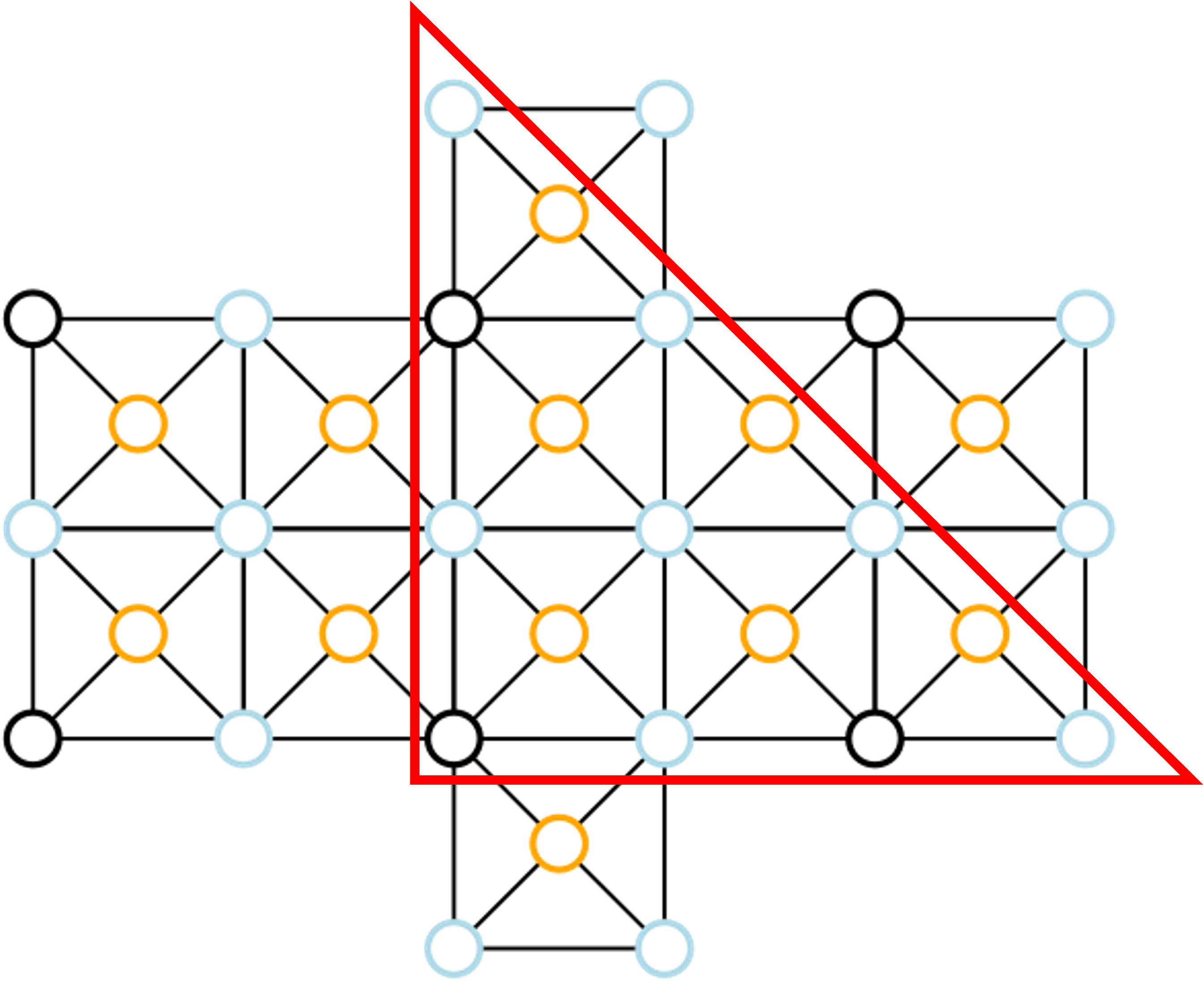}}\hfill\subfloat[New topology of a $P_4$ defined by the dense embedding]{\includegraphics[scale=0.35]{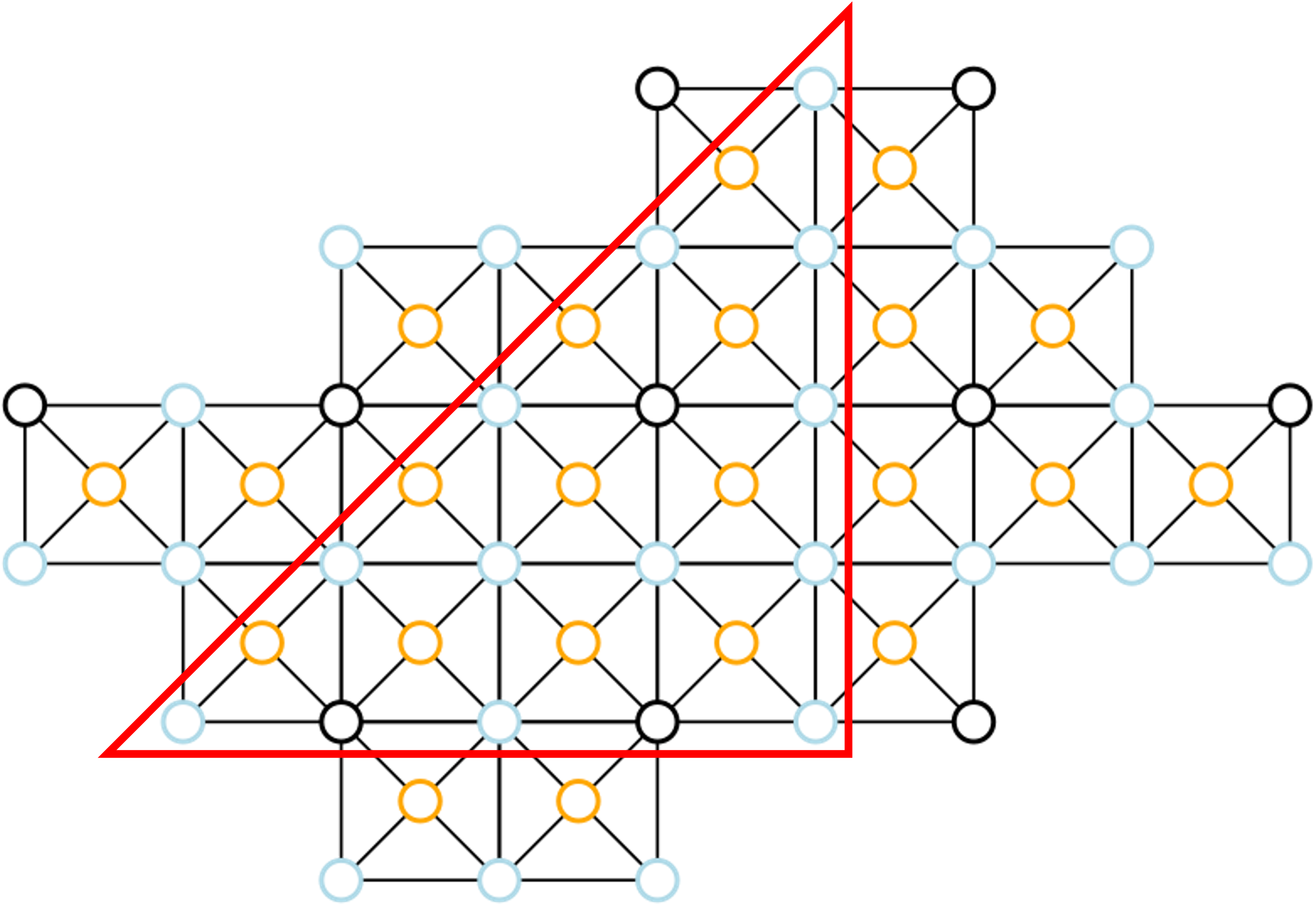}}
    \caption{New topologies of the quantum annealer after considering the chains of spins as parity and auxiliary qubits. The light-blue dots are parity qubits; the black dots are parity qubits where the local field is flipped; and, the orange qubits are the auxiliary qubits that can be used to implement square or triangular plaquettes. The dense embedding results in a larger topology graph because the minimal component, two adjacent plaquettes, can be embedded by using fewer diamonds. Moreover, in the pictures, the largest implementable LHZ triangles are highlighted. We can use these new topologies to implement the LHZ triangle without using any further compilation techniques or, by means of a parity compilation find the right parity compiled problem that fits the topology.}
    \label{fig: new topo}
\end{figure*}

We next implement parity compiled problems by considering the new hardware topology defined by the original and dense embeddings, first described in \cref{sec: parity on dwave}. The new topology hardware graphs created by these embeddings are shown in \cref{fig: new topo}. However, when the quantum hardware is produced it can contain defects. 
In current quantum annealing hardware, not all spins are available to be used, due to hardware constraints or defective manufacturing. 
These defects manifest themselves as missing parity and/or auxiliary qubits in the new topologies. Thus, the new topologies are not entirely available in the hardware since not all the parity and auxiliary qubits are present. Moreover, this new topology with missing qubits changes from quantum annealer to quantum annealer since the defected spins differ. 
To overcome this issue, we can find, either manually or by using a packing heuristic, the biggest implementable LHZ triangle for each topology. Or, we can use a parity compilation method to find the correct parity compiled Hamiltonian that fits the new topology. In this work, we focus on the first case and we identify an LHZ triangle for the original and dense topology for D-Wave's ``Advantage 4.1'' quantum annealer.

To understand how parity compiled problems perform in an open quantum system, we first consider the smallest possible implementation: a single square odd parity plaquette. We encode a single square plaquette as
\begin{equation}\label{eq: square ham}
    H_\square=\left(2\tilde{\sigma'}_z^{l,a} + \tilde{\sigma'}_z^{l,n}+\tilde{\sigma'}_z^{l,e}+\tilde{\sigma'}_z^{l,w}+\tilde{\sigma'}_z^{l,s}\right)^{2}
\end{equation}
and we implement it into a quantum annealer by using the embeddings we developed. 

The performance of QA can be analyzed by considering the percentage of the exact ground state of the embedded Hamiltonian and the outcome wavefunction that can be read from the quantum annealer. Thus, we compute the percentage of samples collected from the annealer that are a global minima of the embedded $H_\square$. 

It is known that the performance of QA depends on different factors~\cite{DWave2009doc} and that a proper choice of parameters influences the results. A feature of quantum annealer processors is the control of the starting time of the evolution of the single spins. By default, spins evolve simultaneously starting at $t=0$ and finishing at $t=T$. However, a custom delay or advance $\Delta t_i$ can be applied to each spin $s_i$--- we call $\Delta t_i$ the annealing offset of the spin $s_i$. Evolving each spin $s_i$ from $t=\Delta t_i$ to $t=T+\Delta t_i$ has been proven to influence the quality of the solutions. The optimal annealing offsets depend on the energy landscape of the Hamiltonian implemented in the quantum annealer and, therefore, it is hard to compute in general~\cite{yarkoni2019boosting}. In~\cite{cattelan2023parallel}, the authors proposed to use the annealing offsets as hyperparameters to be optimized by a classical subroutine. Hence, no prior knowledge of the energy landscape of the Ising Hamiltonian is required. 

Another choice that influences the performance of the algorithm is the annealing time, $T$. Since the coherent time of the spins in the system is smaller than the fastest implementable QA~\cite{ozfidan2020demonstration}, if the system evolves for a longer time we can observe better quality solutions due to coupling with the thermal bath that quenches the excited states in lower energy states~\cite{amin2015searching, raymond2016global}. Therefore, every experiment is conducted for two different annealing times $T=0.5\mu s$ and $T=20 \mu s$ to highlight how the system behaves when it is subject to different thermal effects.

Furthermore, to reduce the effect of noise produced by the electrical leaking from spins to couplers~\cite{pelofske2019optimizing}, we apply a Gauge transformation of the Hamiltonian, called spin-reversal transform, that prevents this effect. Hence, by applying this transformation to the Hamiltonian every $100$ samples we can boost the performance of QA~\cite{DWave2009doc}.

We embed $H_\square$ and we solve it with QA by setting the annealing offsets in three different ways: first, we compute the performance of QA with the default setting, i.e. by letting the spins evolve simultaneously starting from $t=0$; then, we consider a hyperparameter $\theta_i$ for each chain of spins $i$ in the embedding and we set the annealing offsets of the spins in that chain to be $\Delta t_j = \alpha_j\left(1-\theta_i\right)+\beta_j\theta_i$, where $[\alpha_j, \beta_j]$ is the annealing offset range, that is the possible advance or delay of the $j$-th spin in the chain; and eventually, we set the same annealing offset for each spin in a chain $i$ by setting it as $\Delta t_j = {\boldsymbol{\alpha}}_i\left(1-\theta_i\right) + {\boldsymbol{\beta}}_i\theta_i$, where ${\boldsymbol{\alpha}}_i=\max\{\alpha_{i_0}, \ldots, \alpha_{i_n}\}$, ${\boldsymbol{\beta}}_i=\min\{\beta_{i_0}, \ldots, \beta_{i_n}\}$ and where $[\alpha_{i_j}, \beta_{i_j}]$ is the annealing offset range of the $j$-th spin.

\begin{figure*}
    \centering
    \subfloat[Performance of the $H_\square$ for $T=20\mu s$ with different annealing offsets]{\includegraphics[scale=0.6]{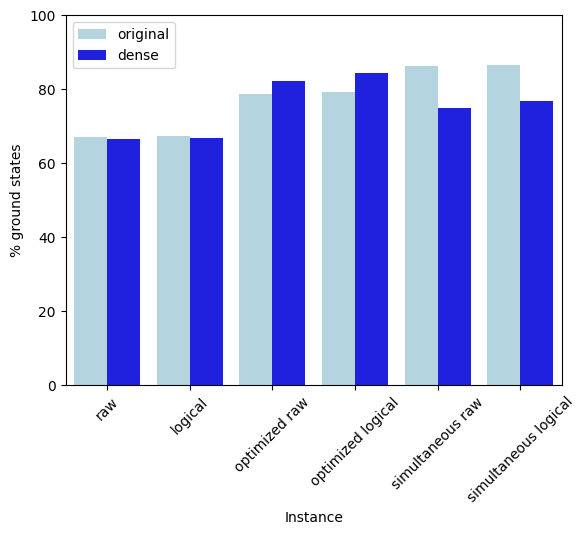}}\hfill\subfloat[Performance of the $H_\square$ for $T=0.5\mu s$ with different annealing offsets]{\includegraphics[scale=0.6]{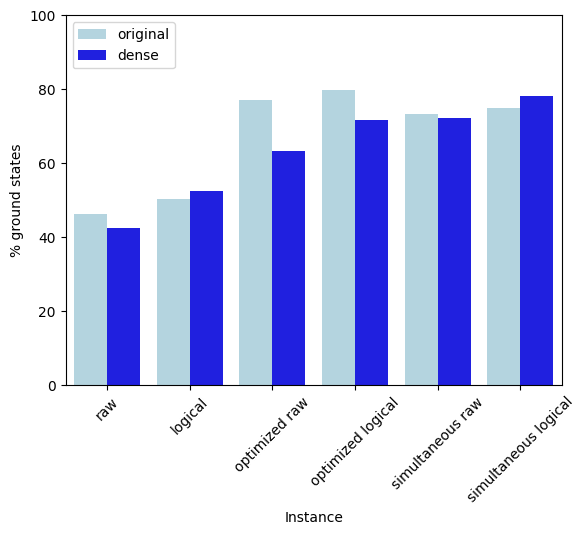}}
        \caption{Percentage of the exact ground states of the embedded Hamiltonian $H_\square$. QA was executed with three different choices of annealing offsets. Every experiment was executed for the original (\textit{light blue}) and dense (\textit{dark blue}) embedding. We collect $10,000$ samples (\textit{raw}) and fix the broken chain by using a majority vote (\textit{logical}) and we compute the ground state by considering the exact ground state of $H_\square$. The terms ``\textit{optimized}'' (``\textit{simultaneous}'') reflect the choice of letting the spins in the chains evolve starting from different (the same) $\Delta t$. Notice that the logical results do not differ significantly from the raw ones. Thus, the chosen chain strength ensures that the value of the chains of spins is preserved in most cases. Furthermore, we stress that even though the size of the system implemented with the original embedding is larger than the dense embedding system ($20$ spins in the original embedding and $10$ for the dense one), the performance is similar. The same trend can be seen in further experiments presented in this work.}
    \label{fig: square performance}
\end{figure*}

Furthermore, after collecting the samples from the annealer, we inspect the values of the spins in each chain and we fix the broken chains by using a majority vote correction, as described in \cref{sec: QA}. The results of these experiments are shown in \cref{fig: square performance}. Notice that fixing the chains of spins in the solutions does not increase significantly the performance of QA. Thus, the number of broken chains is small compared to the number of samples and, therefore, the chain strength used, i.e. the coefficient of the new term added to the Hamiltonian, is large enough to ensure the validity of the chains. However, it is worth mentioning that the number of broken chains is higher for the dense embedding than for the original embedding. This is due to the largest chain length that in the latter is $5$ while in the former is $4$. This is stressed more in the experiments with $T=0.5\mu s$. The chain strength is computed by using the default \texttt{uniform\_torque\_compensation} method~\cite{DWave2009doc}. 

The results of the experiments in \cref{fig: square performance} show that the optimization of the annealing offsets leads to improved performance. Nevertheless, we cannot conclude which method is the most effective since the results depend both on the choice of the annealing time and the embedding used. Furthermore, due to restricted access to the hardware, the experiments are conducted only on a specific sub-system of spins in the quantum annealer. Therefore, even though the optimization of the annealing offsets improves the percentage of ground states sampled from the annealer, we cannot infer any conclusion. However, a general behavior under the effect of the optimization of the ground state could be achieved by considering the average performance of all the plaquettes obtainable with the embeddings. 

\begin{figure*}
    \centering
    \subfloat[Distribution of the different ground states sampled from QA with $T=20\mu s$]{\includegraphics[scale=0.25]{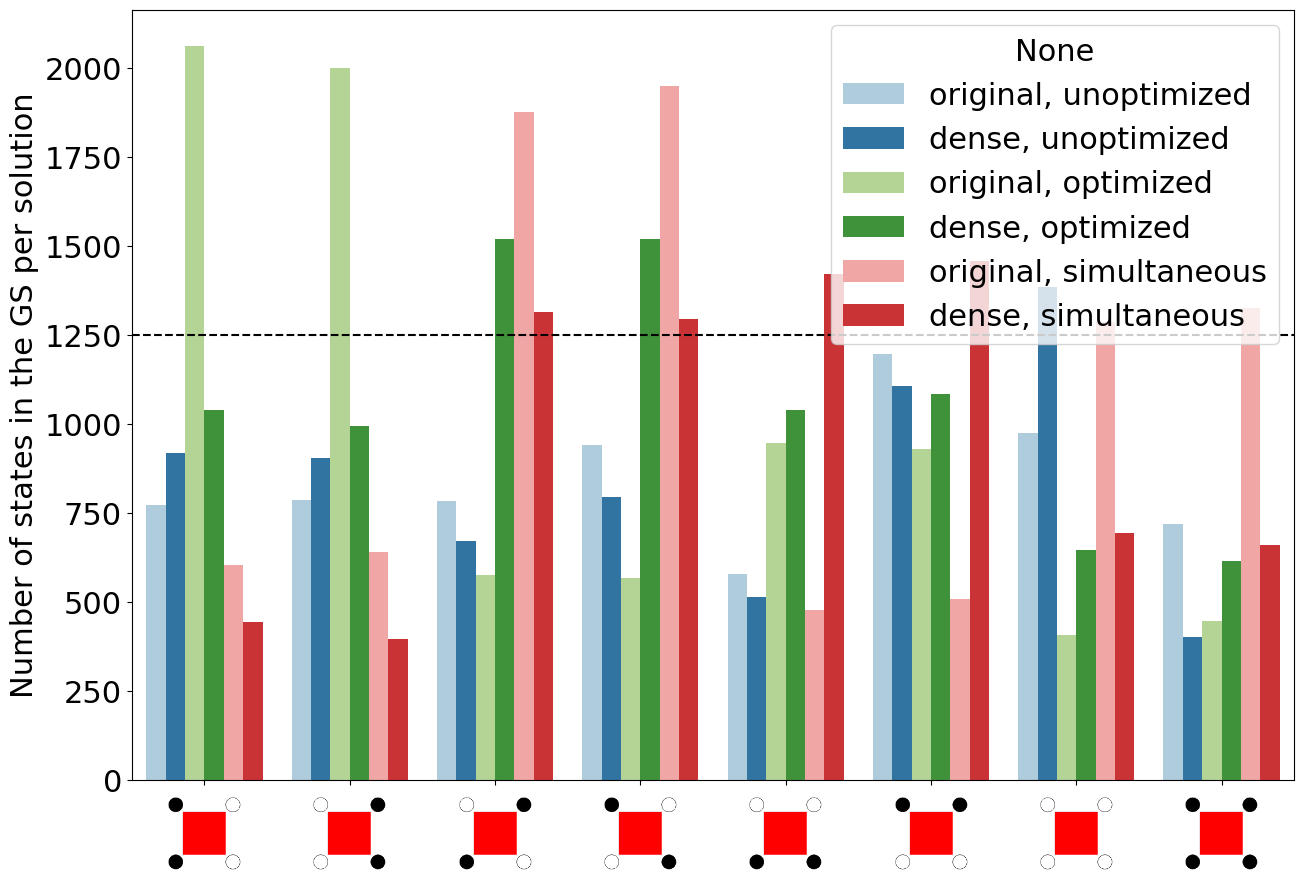}}\hfill\subfloat[Distribution of the different ground states sampled from QA with $T=0.5\mu s$]{\includegraphics[scale=0.25]{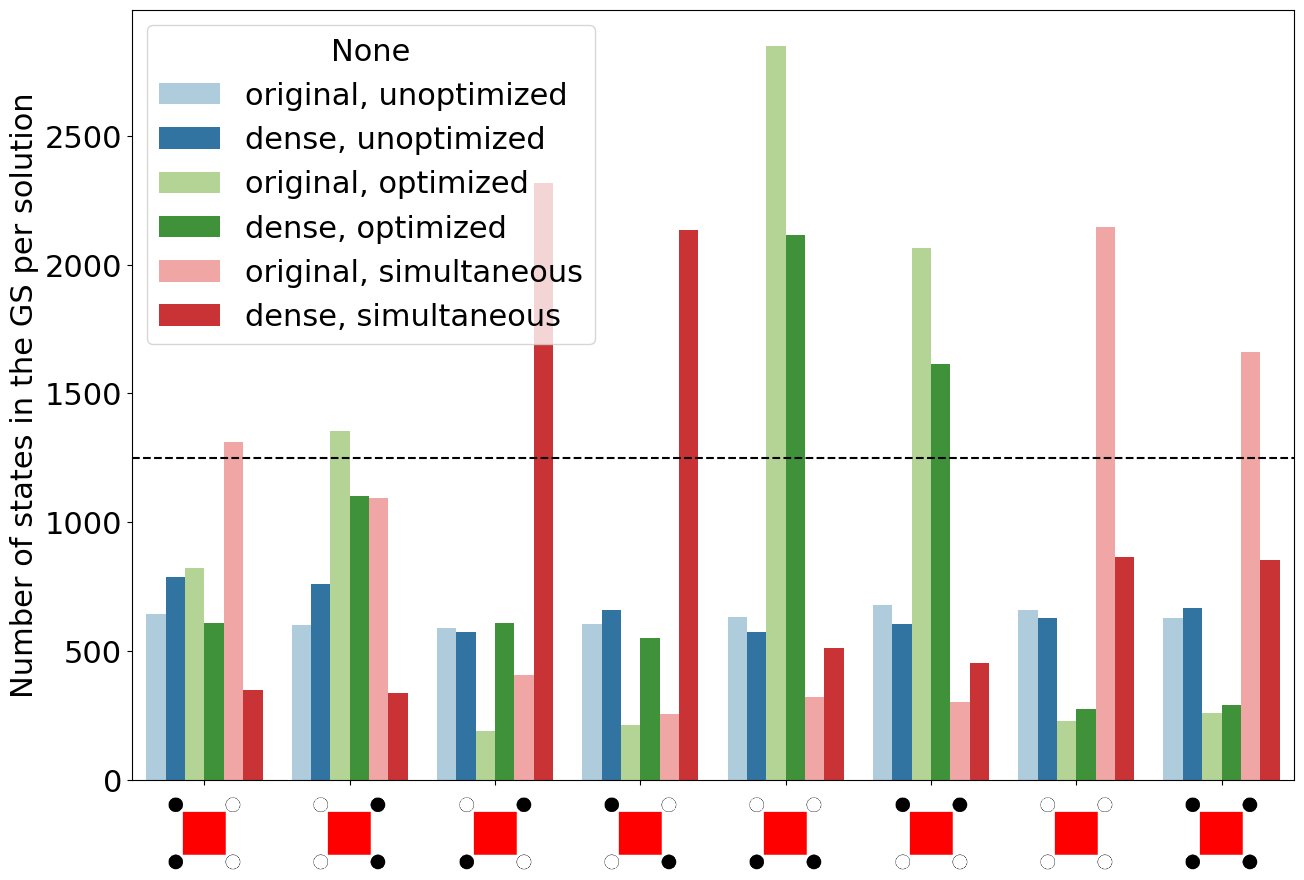}}
    \caption{Distribution of the different ground states collected from the annealer. The colors identify the different optimization of the annealing offsets: every spin begins its evolution with $\Delta t=0$ (\textit{blue}); every spin has its own $\Delta t$ (\textit{green}); and, all the spins in a chain have a unique $\Delta t$ that differs chain to chain (\textit{red}). Every experiment is conducted with the original (\textit{light color}) and dense (\textit{dark color}) embedding. The ticks on the $x$-axis are the different ground states that are the minimum of \cref{eq: square ham}. The theoretical distribution of ground states is uniform, therefore, since we collect $10,000$ samples the ground states should be distributed uniformly and we should count $1250$ of each. Notice that the distribution of the states collected from the annealer after the optimization of the annealing offsets is not uniform. We think that this is due to a combined effect of the classical search algorithm whose goal is to minimize the objective function given without taking into account the physical properties of the system and of the noise that deviates the distribution of states towards specific states. Notice that we can observe the same effect of noise for the distribution of the ground states for the samples collected for \textbf{a)} $T=20\mu s$, since in this case the annealing offsets are not optimized. On the other hand, a distribution more similar to the expected one is shown by the samples collected for \textbf{b)} $t=0.5\mu s$ without optimization of the annealing offsets. The distribution of ground states in this case appears to be closer to the uniform distribution. We think that this behavior can be linked to the time of the annealing. In this case, the period of the anneal is close to the coherent time of the spins and, therefore, the quantum effect might not be influenced by the sources of noise of the hardware too much. Although we have to stress that to confirm this hypothesis, further and deeper studies must be done on the hardware to understand whether the amount of quantum effect can be linked directly to the distribution of ground states. In \cref{fig: distribution comparison} a different analysis of these results is presented.}
    \label{fig: gs distribution}
\end{figure*}

\begin{figure*}
    \centering
    \subfloat[Mean and variance of the distribution of ground states with $T=20\mu s$]{\includegraphics[scale=0.27]{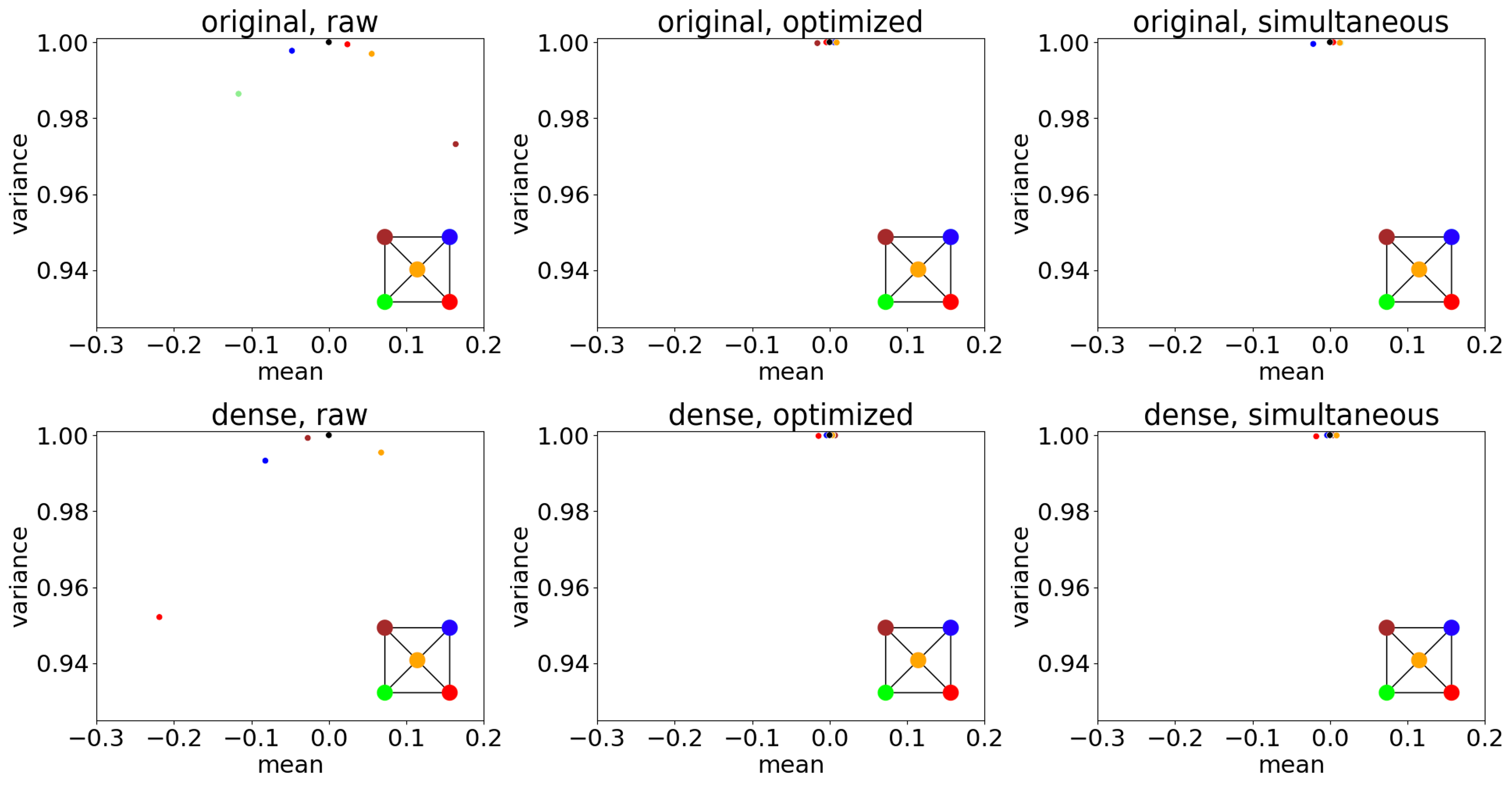}}\hfill\subfloat[Mean and variance of the distribution of ground states with $T=0.5\mu s$]{\includegraphics[scale=0.27]{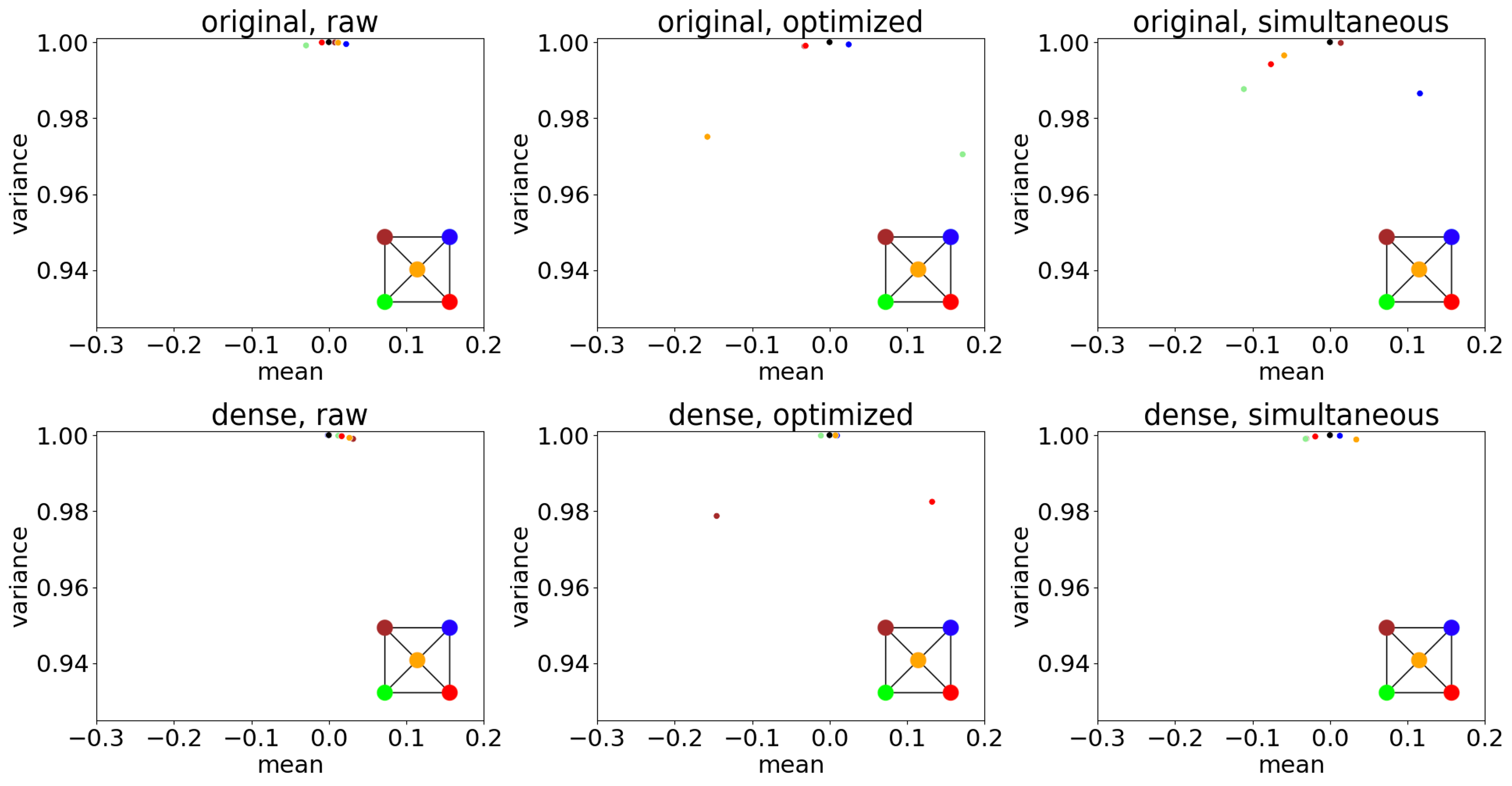}}
    \caption{Mean and variance of the distribution of ground states of the samples collected from the quantum annealer. The mean is the average magnetization of the parity qubits and the auxiliary qubits, which are represented by chains of physical spins in the quantum annealer. The black dot represents the theoretical mean and variance of the distribution of all of the qubits. Each colored dot represents a qubit in the plaquette shown in the lower right corner. Notice that the distribution collected with annealing time $T=0.5\mu s$ and without applying any optimization of the annealing offsets is the closest to the theoretical expectation. Furthermore, we can see that the mean and the variance of the distributions obtained from the samples collected from QA with $T=20\mu s$ are close to the theoretical expectations as well. This is due to the symmetry of the states that appear in the distribution collected from the annealer. We can see that a state appears in the distribution almost with the same frequency as its complementary, i.e. the state whose spins have opposite values. Nevertheless, the distribution of those states is not uniform as shown in \cref{fig: gs distribution}.}
    \label{fig: distribution comparison}
\end{figure*}

Moreover, we can consider the distribution of the valid configurations of the square plaquette, which are the global minima of $H_{\square}$. In a closed quantum system, if the adiabatic condition is satisfied, the theoretical distribution of the valid configurations is uniform. Therefore, in our case, if we collect $10,000$ samples, every valid configuration should appear exactly $1250$ times (within statistical error), since there are $8$ minima. On the other hand, in the samples collected from the annealer, we can observe different results. In \cref{fig: gs distribution}, we can see that the outcome distribution of the ground states from the hardware varies according to the chosen annealing time and optimization method for the annealing offsets. Even though the optimization of the annealing offsets increases the percentage of ground states overall, the properties of the ground state distribution are not considered as a feature in the optimization. Thus, for both $T=0.5\mu s$ and $T=20\mu s$ the ground state distribution is not uniform. The same conclusion can be drawn for the samples obtained from QA with $T=20\mu s$ optimization of the annealing offsets. 
However, notice that even though the distribution of ground states collected from QA with optimized annealing offsets is not uniform, as shown in \cref{fig: distribution comparison}, the variance and the average of the magnetization of the parity qubits, i.e. the chain of physical spins that represent parity qubits, are close to the theoretical expectations. 
We believe this can be traced back to the combined action of noise that biases the samples in the distribution and the optimizer that tries to amplify this effect to improve the expected values of the embedded Hamiltonian. 

Eventually, we can see that with $T=0.5\mu s$, the distribution of the state is close to being uniform. We think this is due to $T=0.5\mu s$ being the fastest evolution we can implement.
Therefore, we expect that the outcome samples have the largest overlap with the theoretical ground state of the Hamiltonian, even though the evolution is carried out when the spins are not coherent anymore. To verify it, since we cannot do state tomography with current quantum annealers we check that the mean and the variance of the distribution of the ground state are close to the theoretical expectations. In \cref{fig: distribution comparison}, the comparison between the theoretical expectation and the experimental results is shown. Although to be able to understand and describe this phenomenon, further studies on the open quantum system should be made which are not included in this analysis due to limited hardware access.

Moreover, even though the system size of the problem embedded by using the original embedding is double the system size of the dense embedding implementation ($20$ spins in the original embedded plaquette and $12$ spins in the dense embedded plaquette, see \cref{fig: embeddings}), the results are similar. This same behavior can be seen when we consider larger parity compiled problems. 
In all the cases, the number of spins used when embedding the parity compilation with the original embedding was larger than the number of spins used by the dense embedding and the results do not differ significantly. 

\begin{figure*}
    \centering
    \subfloat[QA with $T=20\mu s$]{\includegraphics[scale=0.25]{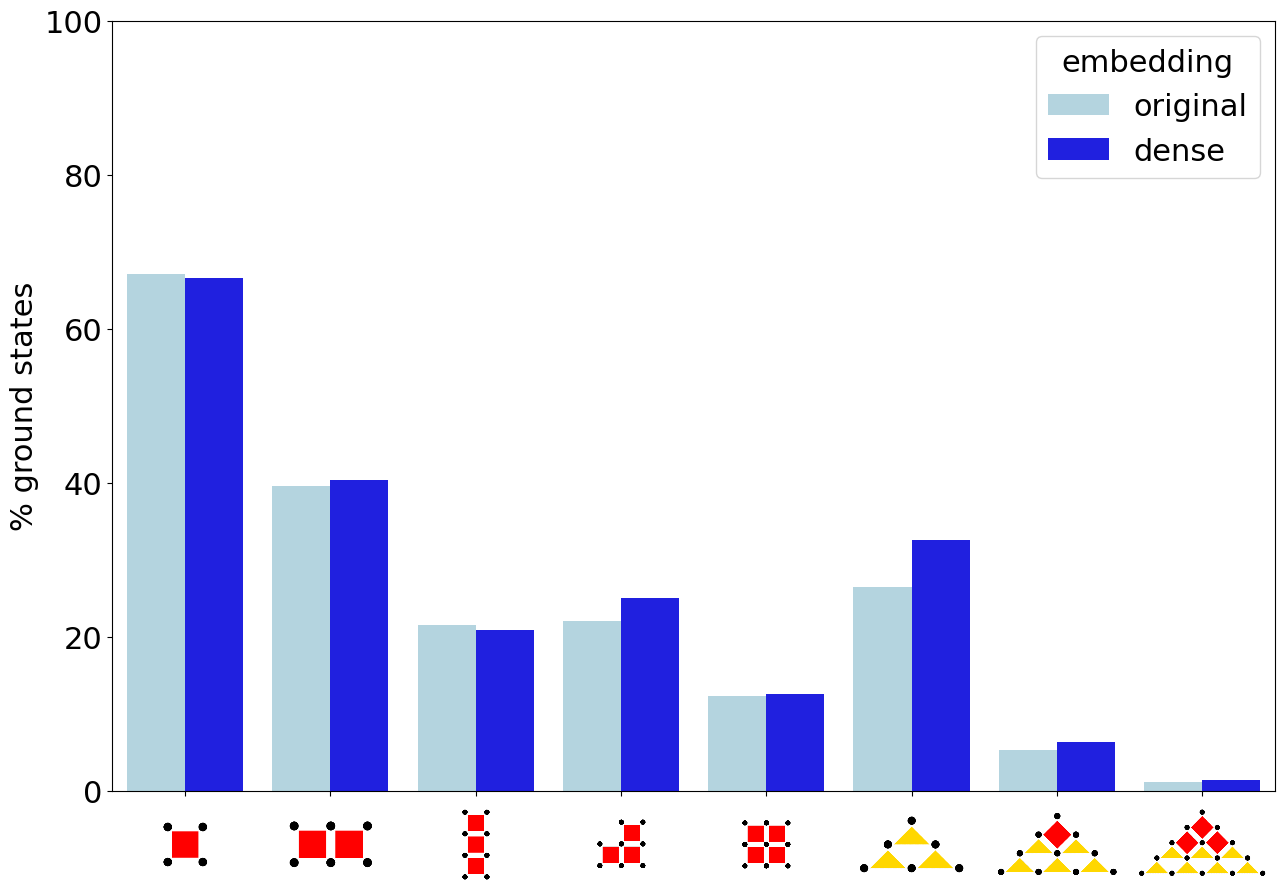}}\hfill\subfloat[QA with $T=0.5\mu s$]{\includegraphics[scale=0.25]{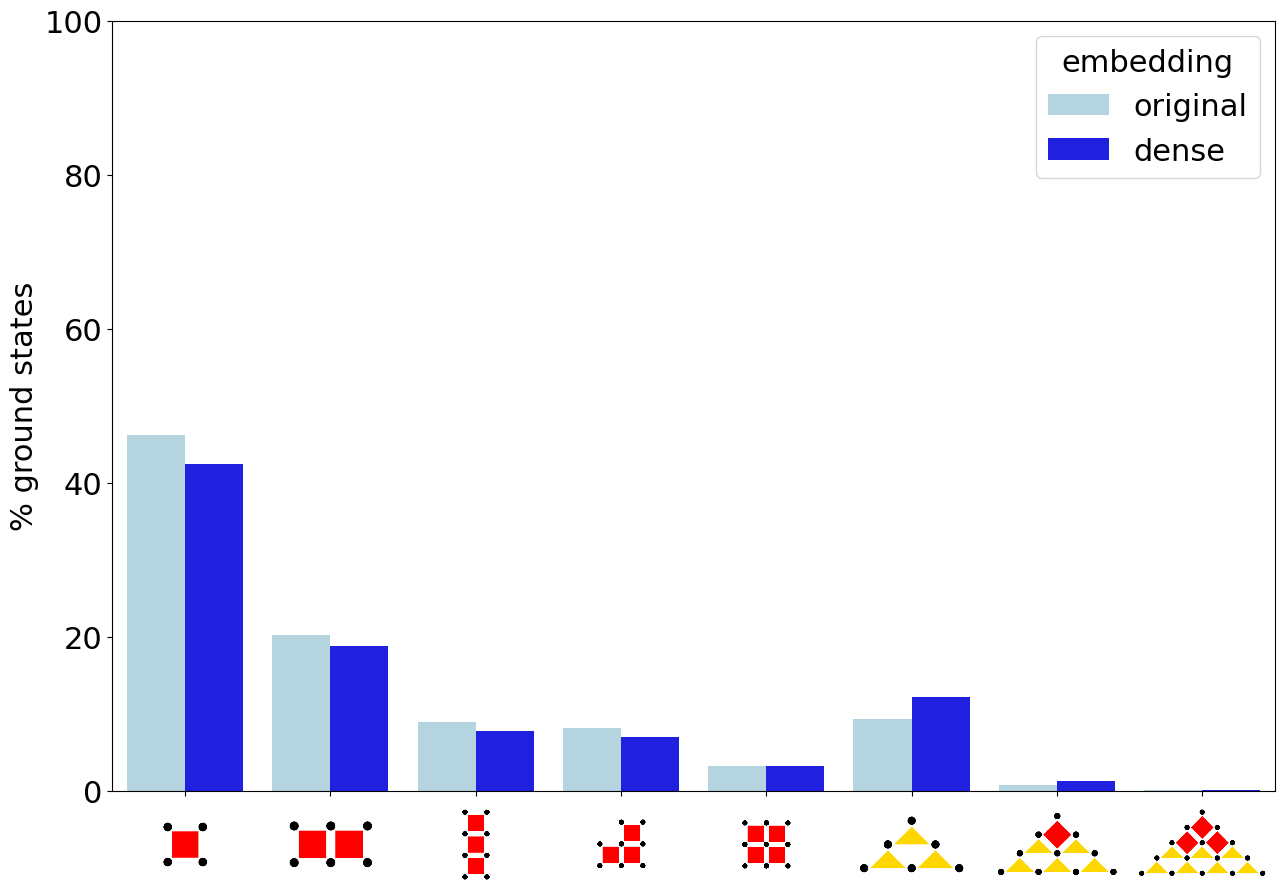}}
    \caption{Performance of QA on different parity compiled problems and annealing time. We choose all the possible combinations of parity compiled problems up to rotations with many plaquettes from $1$ to $3$ with only square plaquettes, a square of $4$ square plaquettes and LHZ triangle with many qubits in the basis from $3$ to $5$. We embed the problems into the quantum annealer by using the two embeddings we developed: original (\textit{light blue}) and dense (\textit{dark blue}). To be able to measure the ground states of the problems we decided to implement the most degenerate problem by setting all the local fields to $0$. We solve every instance with annealing time $T=0.5\mu s$ and $T=20\mu s$. For each experiment, we collect $10,000$ samples. As already shown in ~\cite{amin2015searching}, setting a longer annealing time leads to improved results. Notice that the performance of QA decreases by increasing the number of parity qubits. However, we can see that the two embeddings perform similarly despite the number of spins used in the embedded model almost double for the original embedding compared with the dense one. }
    \label{fig: empty plaquettes}
\end{figure*}

In \cref{fig: empty plaquettes}, we compare the performance of QA across various geometries and parity plaquette combinations. Here we set all the local fields of the parity qubits to $0$. 
Also in this case we stress that the percentage of ground states collected from QA is similar independent of the choice of the embedding used. As noted above, we think that this can be explained by considering the chain length of the chains used. Whereas the original embedding consists of chains of length $4$, the dense embedding has chains length of at most $5$. This can influence the performance of the algorithm as is noticed in~\cite{boothby2016fast}. Moreover, we stress that the geometry of the parity compiled problem matters and that parity compiled problem composed by similar plaquettes returns similar results.

Finally, we complete the analysis of the minimum spectral gap points by considering the simulation presented in \cref{sec: simulation} and identifying for each instance the position of the minimum spectral gap point in the evolution. The position of the minimum spectral gap point is crucial because if it occurs earlier in the anneal the system will be driven away from the ground state due to noise and coupling with the environment.
On the other hand, if it occurs later in the anneal the system will not be able to repopulate the low states from the excitations that occurred before the minimum spectral gap~\cite{marshall2019power, dickson2013thermally}. 

\begin{figure*}
    \centering
    \subfloat[(2, 1, 1)]{\includegraphics[scale=0.37]{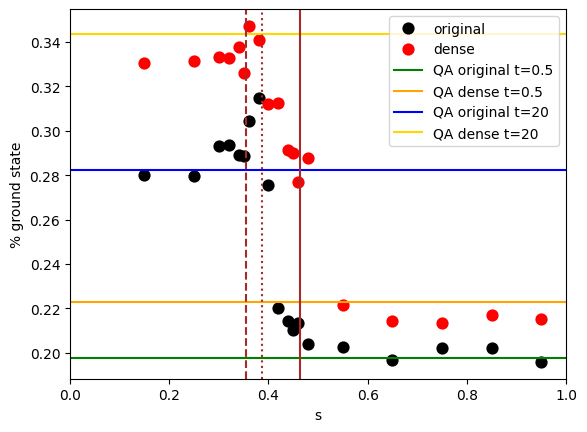}}\hfill\subfloat[(3, 1, 1)]{\includegraphics[scale=0.37]{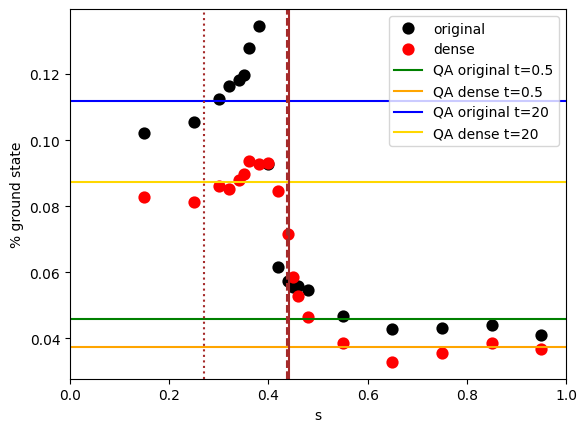}}\hfill\subfloat[(3, 1, 2)]{\includegraphics[scale=0.37]{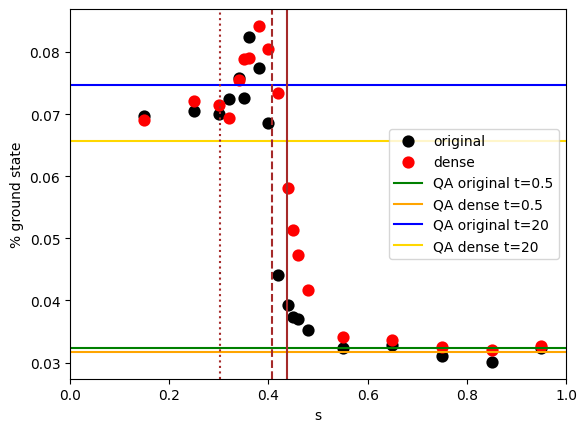}}

    \subfloat[(4, 1, 1)]{\includegraphics[scale=0.37]{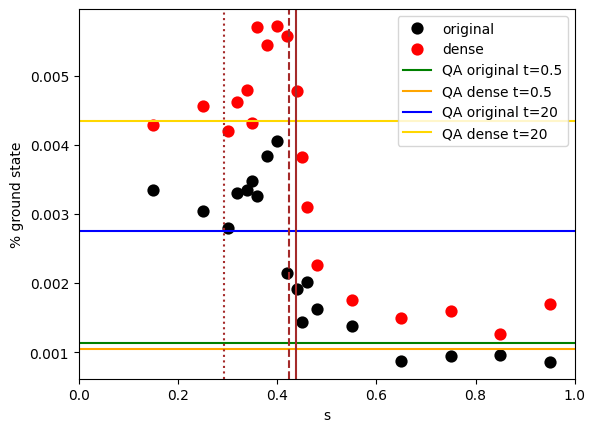}}\hfill\subfloat[(4, 1, 2)]{\includegraphics[scale=0.37]{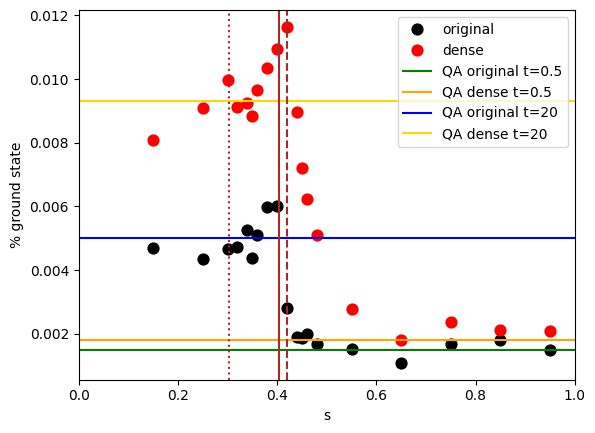}}\hfill\subfloat[(4, 1, 3)]{\includegraphics[scale=0.37]{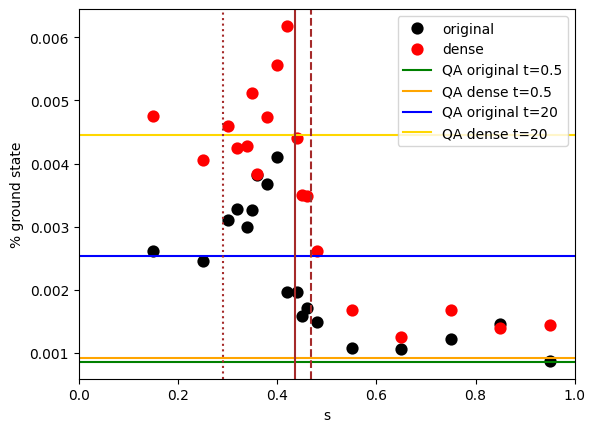}}

    \subfloat[(4, 2, 1)]{\includegraphics[scale=0.37]{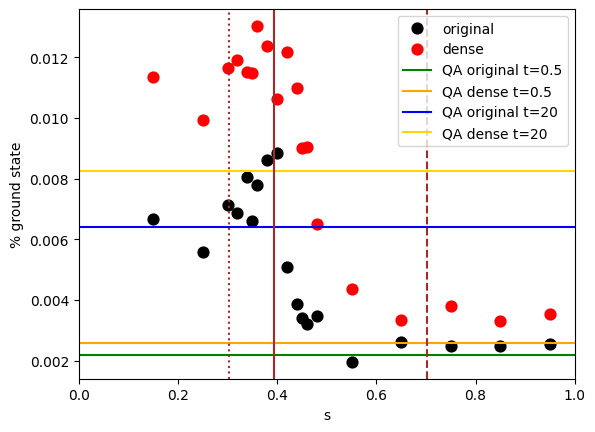}}\hfill\subfloat[(4, 2, 1)]{\includegraphics[scale=0.37]{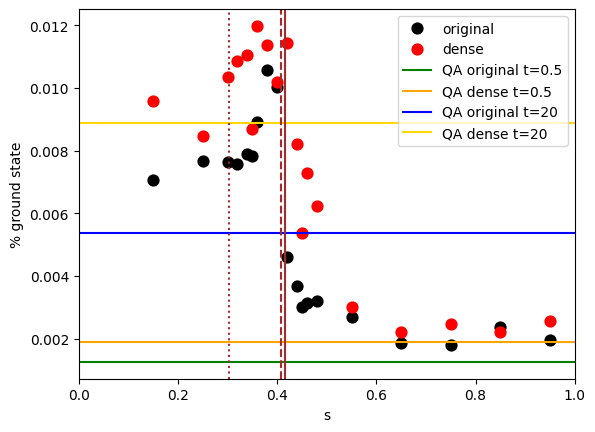}}\hfill\subfloat[(4, 2, 1)]{\includegraphics[scale=0.37]{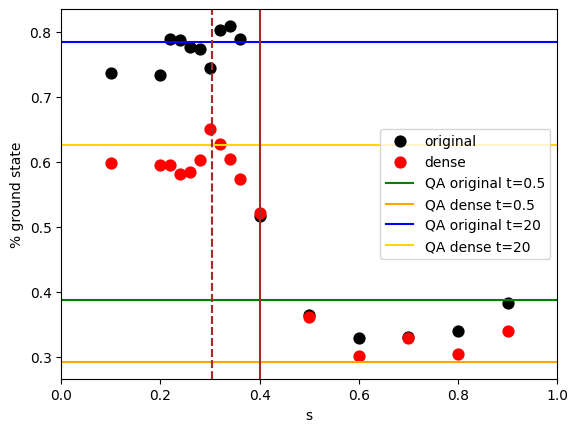}}
    \caption{Results of the pausing technique explained in \cref{subsec: exp}. The solid horizontal lines are the references for the QA performance if performed without any modification of the annealing path, with $T=0.5\mu s$ and $T=20\mu s$, and by using the different proposed embeddings. The vertical straight lines indicate the location of the minimum spectral gap point obtained by the simulation of the logical frustrated Ising Hamiltonian (\textit{solid line}), of the multi-body LHZ (\textit{dotted line}) and of the $2$-body LHZ as described in \cref{fig: qubo LHZ} with interactions tuned to implement the multi-car paint shop problem (\textit{dashed line}). The dots are the average performance of QA when the pause happens at the specific $s$. Those points are obtained by execution of QA solving the embedded $2$-body LHZ triangle that implements different multi-car paint-shop problems. $50.000$ samples are collected to generate each point. The original (\textit{black dots}) and dense (\textit{red dots}) embeddings are used to show differences between them. Notice that in every plot a peak of the performance can be observed. This peak is the result of pausing close to the minimum spectral gap location. This technique allows us to identify the position of the minimum spectral gap point of the embedded Hamiltonian, which cannot be computed classically due to the large number of spins involved. We can see that in most instances all the minimum spectral gap points are close to each other. Every instance is associated with a tuple in the caption to give a compact label. For instance, in the subplot \textbf{a)} the instance with $2$ cars in the problem, $1$ group of cars with the same model ($|G_0|=2$) and only $1$ car to be painted in black ($k_0=1$) is shown. The last three instances are described by the same tuple, but the groups of cars are different and identify three different non-trivial instances.}
    \label{fig: pausing}
\end{figure*}

As the problem size increases, the minimum spectral gap point of the embedded Hamiltonian in the quantum annealer cannot be efficiently simulated classically because the number of spins required to solve the problem with quantum annealers quickly exceeds the problem size we can simulate with our classical hardware.
Therefore, to understand the position of the minimum spectral gap of the embedded Hamiltonian we need to use experimental data. As already noted in~\cite{marshall2019power}, we can experience a boost in the performance of QA if we pause the evolution of the QA close to the minimum spectral gap point. Thus, by inspecting the performance of different executions of QA with pauses in different points we can obtain an approximate location where the minimum spectral gap point takes place. 
Moreover, to be able to have a better approximation of the position of the minimum spectral gap point we propose a different approach based on the pausing technique. It is known that crossing the minimum spectral gap point slowly during the evolution of the QA system boosts the performance of the algorithm. Hence, we can modify the rate of the annealing schedule such that the evolution leads to better performance if we pause before the minimum spectral gap point. To achieve this, we can start the evolution by setting the rate of the annealing schedule to be as fast as possible, then pause the system and eventually evolve it slowly until the end. For instance, if we want to pause the evolution at $s^*$, we customize the annealing schedule to cross these points: starting at $s=0$ and $t=0$; then, evolving as fast as possible to $s=s^*$ and $t=0.5s^*$; pausing for $10\mu s$; and evolving slowly to end the protocol at $s=1$ and $t=T=20\mu s$. Notice that by applying this annealing schedule if we pause before the minimum spectral gap point we cross it slowly, whereas if we pause later, the minimum spectral gap point is crossed with the highest rate implementable. Therefore, the performance of QA is better for $s^*$ smaller than the minimum spectral gap point, with a possible peak if we pause close to it, and worse for $s^*$ larger than the minimum spectral gap point. We can observe this behavior in \cref{fig: pausing}. The vertical solid lines represent the minimum spectral gap points for the problems implemented as frustrated Ising chain, multi-body LHZ and $2$-body LHZ. The dots are the average probability of measuring the ground state and their abscissa represents the $s^*$ where the pause takes place. We can observe a peak in the results. Those improved points identify the position of the minimum spectral gap location of the embedded Hamiltonian. Notice that all four minimum spectral gap points are close (in an interval of length $0.2$) to each other for most of the instances. Thus, despite the redundant encoding of the Hamiltonian through the parity mapping, the location of the minimum spectral gap point remains unaffected across different implementations for most of the instances analyzed.

\section{Discussion}

In this work, we presented an extension of the parity mapping that embeds combinatorial optimization problems as a parity compiled Hamiltonian onto quantum annealers. We showed how to build two different embeddings to implement the parity compiled Hamiltonians with terms of at most quadratic order in hardware.

Using the multi-car paint shop problem as a test-bed, we numerically simulated quantum annealing and compared the performance, evaluated as spectral gap size, of equivalent instances compiled using different implementations, namely: logical, multi-body LHZ and 2-body LHZ. We showed that even though the redundant implementations bring a larger overhead of spins, the value of the minimum spectral gap does not necessarily shrink and it can become larger for some instances.

By implementing parity compiled problems using both the original and dense embeddings experimentally on current D-wave annealing hardware, we first compared the performance, ground state distribution, and ground state variance of a single square plaquette Hamiltonian at different annealing times over several optimization schemes. Here we found that shorter annealing times (without any advance or delay to the evolution of the physical spins) bring a distribution closer to the theoretical expectation. 

We also compared the performance for the original and dense embeddings across several geometries of parity embedded Hamiltonians for annealing times of $20 \mu s$ and $0.5 \mu s$. Especially, we pointed out that different embeddings of the same parity compiled problem behave similarly despite the different number of physical spins involved in the implementation onto the hardware quantum system. 
Finally, we experimentally approximated the positions of the minimum spectral gap of several instances of the multi-car paint shop problem by implementing the Hamiltonians with the original and dense embedding. We compared the experimental gap positions with the positions of the minimum spectral gap in the logical, multi-body LHZ, and $2$-body LHZ implementations obtained by the simulation, showing that the minimum spectral gap points take place roughly at the same time for most of the instances. Therefore, we see that despite its larger spin overhead, the extension of the parity map presented preserves the physical properties of the logical system.

Even though we can conclude that the performance of QA decreases with the size of the problem implemented, it is not clear whether a larger implementation of the problem as a parity compiled Hamiltonian could lead to improved results. 
The fixed maximum chain length and the upper bound on the number of spins required to embed an arbitrary problem suggests that the performance of QA for parity compiled Hamiltonian could be comparable, if not better, to the performance of QA for minor embedded problems. 
However, to be able to answer this question, a larger quantum annealer would need to be used. This is because the largest problem size embeddable (using the proposed parity mapping) on the largest currently available quantum annealer is still too small to compete against minor embedding techniques.

In future works, we will focus on finding better and more efficient embeddings to reduce the number of spins and limit the chain length. Furthermore, we will investigate the performance of general parity compiled problems implemented with a heuristic into the new topologies presented to investigate whether different compilation techniques can further improve the performance. The work also motivates the experimental implementation of $4$-body couplers to directly implement the LHZ architecture. 

\section{Acknowledgements}

This research was funded in part, by the Austrian Science Fund (FWF) SFB BeyondC Project No. F7108-N38, START grant under Project No. Y1067-N27 and I 6011. For the purpose of open access, the author has applied a CC BY public copyright licence to any Author Accepted Manuscript version arising from this submission. This project was funded within the QuantERA II Programme that has received funding from the European Union's Horizon 2020 research and innovation programme under Grant Agreement No. 101017733.
The computational results presented here have been achieved in part using the LEO HPC infrastructure of the University of Innsbruck.

\newpage
\bibliography{lib}%

\end{document}